\documentclass[fleqn]{2017SCGE}
\usepackage{graphicx}

\usepackage{mwe,tikz,ulem}
\usepackage[percent]{overpic}
\usepackage{graphicx}
\usepackage{xspace}
\usepackage[numbers]{natbib}
\usepackage[utf8x]{inputenc}

\usepackage[gen]{eurosym}
\usepackage{microtype}
\usepackage{color}
\usepackage{txfonts}
\usepackage{makeidx}
\usepackage[columns=1]{idxlayout}

\begin{document}
\ensubject{subject}

\ArticleType{Article}
\SpecialTopic{SPECIAL TOPIC: }
\Year{2017}
\Month{January}
\Vol{60}
\No{1}
\DOI{10.1007/s11432-016-0037-0}
\ArtNo{000000}

\title{The enhanced X-ray Timing and Polarimetry mission- \textit{eXTP}}{the eXTP mission}


\author[1]{Shuang-Nan Zhang}{{zhangsn@ihep.ac.cn}}
\author[1,2]{Andrea Santangelo}{{andrea.santangelo@uni-tuebingen.de}}
\author[3,4]{Marco Feroci}{{marco.feroci@iaps.inaf.it}}
\author[1]{Yupeng Xu}{{xuyp@ihep.ac.cn}}
\author[1]{Fangjun Lu}{}
\author[1]{\\Yong Chen}{}
\author[5]{Hua Feng}{}
\author[1]{Shu Zhang}{}
\author[36]{S{\o}ren Brandt}{}
\author[12,13]{Margarita Hernanz}{}
% ...

\AuthorMark{Zhang S.N., Santangelo A., Feroci M., Xu Y. }


\AuthorCitation{Zhang S.N., Santangelo A., Feroci M., Xu Y., et al.}


\address[1]{Key Laboratory for Particle Astrophysics, Institute of High Energy Physics, Beijing 100049, China}
\address[2]{Institut f\"{u}r Astronomie und Astrophysik, Eberhard Karls Universit\"{a}t, 72076 T\"{u}bingen, Germany}
\address[3]{INAF -- Istituto di Astrofisica e Planetologia Spaziali, Via Fosso del Cavaliere 100, I-00133 Roma, Italy} 
\address[4]{INFN -- Roma Tor Vergata, Via della Ricerca Scientifica 1, I-00133 Roma, Italy} 
\address[5]{Department of Engineering Physics and Center for Astrophysics, Tsinghua University, Beijing 100084, China}
% ...
\address[12]{Institute of Space Sciences (ICE, CSIC), 08193 Cerdanyola del Vall\`es (Barcelona), Spain}
\address[13]{Institut d'Estudis Espacials de Catalunya(IEEC), 08034 Barcelona, Spain}
% ...
\address[36]{DTU, Building 327, DK-2800 Kongens, Lyngby, Denmark}
% ...

\abstract{In this paper we present the enhanced X-ray Timing and Polarimetry mission. eXTP is a space science mission designed to study fundamental physics under extreme conditions of density, gravity and magnetism. The mission aims at determining the equation of state of matter at supra-nuclear density, measuring effects of QED, and understanding the dynamics of matter in strong-field gravity. In addition to investigating fundamental physics, eXTP will be a very powerful observatory for astrophysics that will provide observations of unprecedented quality on a variety of galactic and extragalactic objects. In particular, its wide field monitoring capabilities will be highly instrumental to detect the electro-magnetic counterparts of gravitational wave sources. The paper provides a detailed description of: 1) The technological and technical aspects, and the expected performance of the instruments of the scientific payload; 2) The elements and functions of the mission, from the spacecraft to the ground segment.}



\maketitle


\begin{multicols}{2}
\section{Introduction}\label{sec:Introduction}

\textit{The enhanced X--ray Timing and Polarimetry mission} -- eXTP is a scientific space mission designed to study the state of matter under extreme conditions of density, gravity and magnetism \cite{Zhang2016}. Primary goals are the determination of the equation of state of matter at supra-nuclear density, the measurement of QED effects in the radiation emerging from highly magnetized stars, and the study of matter dynamics in the strong-field regime of gravity. The matter inside neutron stars (NSs), the space-time close to Black Holes (BHs), and the extremely magnetized vacuum close to magnetars are among the uncharted territories of fundamental physics. The eXTP mission will revolutionize these areas of  fundamental research by high precision X-ray measurements of NSs across the magnetic field scale and BHs across the mass scale. 
In addition to investigating fundamental physics, eXTP will be a very powerful observatory for astrophysics, which will provide observations of unprecedented quality on a variety of galactic and extragalactic objects. In particular, its wide field monitoring capabilities will be highly instrumental in identifying the electro-magnetic counterparts of gravitational wave sources.

The eXTP science case is described in four papers addressing the three main science objectives of the mission and the observatory science \cite{Watts2018, Santangel2018, DeRosa2018, Bozzo2018}. They are included in this special issue.  

\begin{figure}[H]
\centering
\includegraphics[width=0.9\columnwidth, angle=0]{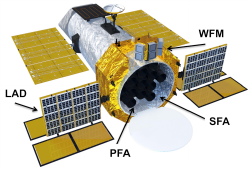}
\caption{Artistic view of the eXTP satellite. The science payload consists of four instruments: the focused SFA and PFA telescopes arrays, the large area instrument LAD, and the WFM to monitor a large fraction of the sky.}
\label{Fig:eXTP_satellite}
\end{figure}

\begin{figure}[H]
\centering
\includegraphics[width=1.0\columnwidth]{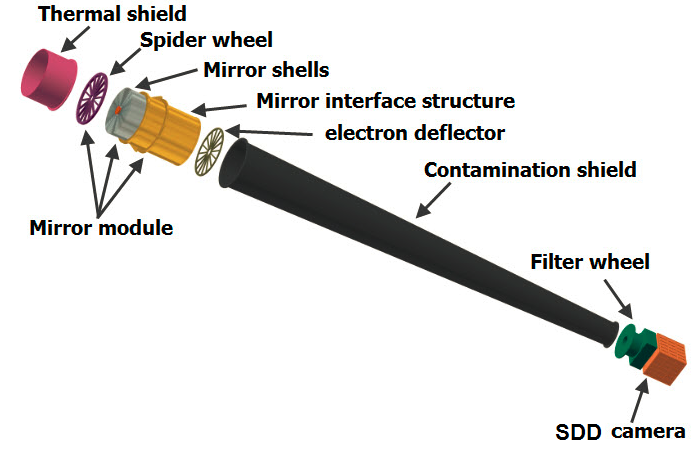}
\caption{The schematic structure of one of the SFA telescopes.}
\label{fig_SFA_ske}
\end{figure}

The eXTP mission is an enhanced version of the Chinese \textit{X-ray Timing and Polarimetry mission} \cite{XTP}, which in 2011 was selected and funded for a Phase 0/A study as one of the background concept missions in the Strategic Priority Space Science Program of the Chinese Academy of Sciences (CAS). Also in 2011, the \textit{Large Observatory for Timing } (LOFT) mission concept \cite{Feroci2012, Feroci2014} was selected for an assessment study in the context of the ESA's Announcement of Opportunity for the third of the medium size missions (M3) foreseen in the framework of the Agency's Cosmic Vision programme. The LOFT study was carried out in 2011-2014 by a consortium of European institutes. Eventually the exo-planetary mission PLATO, considered more well-timed, was selected as ESA’s Cosmic Vision M3 mission. Following this, in 2015, the European LOFT consortium and the Chinese team merged the LOFT and XTP mission concepts, thus starting the eXTP project. 

The eXTP International consortium is led by the Institute of High Energy Physics (IHEP) of CAS, and includes other major institutions of CAS, several Chinese Universities, and institutions from ten European countries. Other international partners participate in eXTP as well. Overall more than 200 scientists in over 100 institutions from about 20 countries are members of the consortium. The mission has recently started an extended phase A study that is expected to be completed by the end of 2018 in view of a launch around 2025.

In this paper we present the technological and technical aspects, and the expected performance of the instruments of the scientific payload, as well as the main elements and functions of the mission. 


\section{The scientific payload}\label{sec:2}

An artistic view of the current design of the eXTP satellite is shown in Figure~\ref{Fig:eXTP_satellite}. 
The scientific payload of the mission consists of four main instruments: the spectroscopic focusing array (SFA, Section \ref{sec:SFA}), the large area detector (LAD, Section \ref{sec:LAD}), the polarimetry focusing array (PFA, Section \ref{sec:PFA}), and the wide field monitor (WFM, Section \ref{sec:WFM}).
\subsection{Spectroscopic Focusing Array - SFA}\label{sec:SFA}
The SFA consists of an array of 9 identical Wolter-I grazing-incidence X-ray telescopes, and is mainly used for spectral and timing observations in the energy range 0.5-10\,keV. Each telescope consists of the thermal shield, the mirror module, the electron deflector, the filter wheel, and the focal plane camera. A schematic view of one of the SFA telescopes is shown in Figure~\ref{fig_SFA_ske}. 
The SFA total effective area is expected to be larger than $\sim7400$\,cm$^2$ at 2\,keV, and the field of view (FoV) is 12\,arcmin in diameter. The SFA uses silicon drift detectors (SDDs) as focal plane detectors. Each telescope includes a 19-cell SDD array, whose energy resolution is less than 180\,eV at 6\,keV. The time resolution is 10\,$\mu$s, and the dead time is expected to be less than 5\% at 1\,Crab. The angular resolution is required to be less than 1\,arcmin (HEW).
To meet the requirements, the working temperature of the mirror modules has to be stable at 20 $\pm$ 2\,$^{\circ}$C. A thermal shield is introduced to keep the temperature as steady as possible. An electron deflector is installed at the bottom of the mirror module to reduce the particle background.


\subsubsection{Optics}

Since they are based on very similar concepts in this section we discuss the optics of both the SFA and PFA. Thirteen X-ray grazing-incidence Wolter I (parabola + hyperbola) optics modules will be implemented onboard eXTP. The SFA includes nine optics modules while four optics modules are designed for the PFA. The expected collecting area of each telescope is A$_{optics}$ $\gtrapprox900$ cm$^2$ at 2 keV, and A$_{optics}$ $\gtrapprox550$ cm$^2$ at 6 keV. The FoV is 12 arcmin (in diameter). These figures already meet the requirement of $\sim 820$ cm$^2$ at 2 keV\footnote{All requirements are being consolidated in the context of the extended phase A study.}. The requirement for the angular resolution of the PFA optics is $<30$ arcsec (HEW, goal is 15 arcsec), more demanding than the 1 arcmin (HEW) angular resolution required for the SFA optics. For both types of telescopes the focal length reference value is 525 cm. The maximum diameter of the mirror shells, allowed by the envelope for a single mirror of less than 600\,mm in diameter,  is about 510 mm. The weight of the optics module is expected to be $\sim$50~kg for the SFA and $\sim$100~kg for the PFA.  In Table \ref{tab1} we list the main parameters of the SFA and PFA optics. 
\begin{table}[H]
\begin{center}
\caption{The main parameters of the SFA and PFA baseline optics.}
\label{tab1}
\footnotesize
\begin{tabular}{l|l}
\bottomrule
\textbf{Parameters} & \textbf{For one telescope} \\
\hline
Focal length & 5.25\,m \\
Aperture & $\le$510\,mm (diameter) \\
Envelope & $\le$600\,mm (diameter) \\
Collecting area & $\ge$900\,cm$^2$@2\,keV \\
Collective area req. & $\sim 820$\,cm$^2$@2\,keV \\
 & $\ge$550\,cm$^2$@6\,keV \\
Energy range & 0.5--10\,keV \\
Field of view & $\ge$12\,arcmin \\
Angular resolution SFA & 1\,arcmin (HPD)\\
Angular resolution PFA & 30'' (goal 15'', HPD) \\
Working temperature & 20 $\pm$ 2\,$^{\circ}$C\\
Weight per module SFA & $\sim 50$\,kg\\
Weight per module PFA & $\sim 100$\,kg\\
\bottomrule
\end{tabular} 
\end{center}
\end{table}
\textit{Nickel electroforming approach.}
The baseline technique implemented for both the SFA and PFA is Nickel replication, developed in Europe by INAF in collaboration with the Media Lario company. Nickel replication has already been successfully used for high throughput X-ray telescopes with good angular resolution, such as BeppoSAX \cite{Citterio}, XMM-Newton \cite{Chambure} and Swift \cite{Burrows}. It is also the technology adopted for the eROSITA  \cite{Predhel} and ART-X telescopes onboard the Spectrum X-GAMMA Mission \cite{Pavlinsky}, and the IXPE \cite{Ramsey} polarimetric mission. The Nickel approach is also the baseline for the realization of the mirrors of the FXT telescope aboard the Einstein Probe Chinese mission \cite{Yuan}. For each mirror shell a mandrel with mirror negative profile (parabola + hyperbola) will be fabricated. The mandrel is made of Aluminum and electrochemically coated with a thin layer (0.100 mm) of Kanigen, a material particularly suitable to endure superpolishing by lapping. A layer of Gold is deposited on the superpolished mandrel surface and the successive deposition is performed by electroforming of the Nickel external substrate. The mirror is then separated from the mandrel by cooling it, exploiting the large difference of the thermal expansion coefficients between the Aluminum of the mandrels and the Nickel of the shells. An overcoating of Carbon is deposited onto the Gold surface to increase the X--ray reflectivity at energies $\leq$ 5 keV \cite{Pareschi}.

\begin{figure}[H]
\centering
  \includegraphics[width=0.5\textwidth]{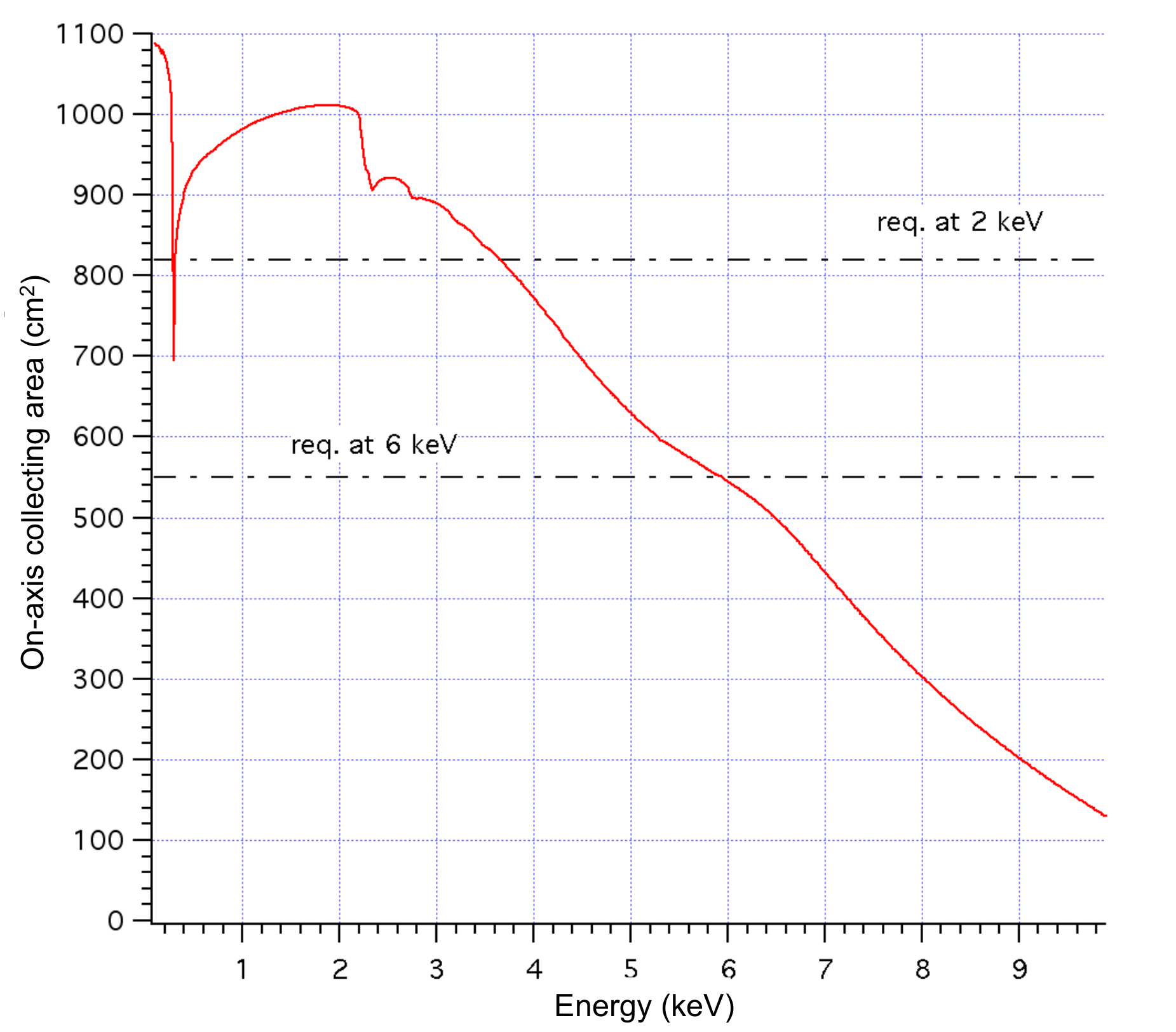}
\caption{On--axis collecting area expected for the eXTP X-ray baseline optics based on Nickel electroforming (single modules). The curve is representative for both the stiff (PFA) and soft (SFA) configurations. The current requirements at 2 and 6 keV are shown. While the requirements are currently being consolidated, we observe that the expected performance already largely meet the requirements.}
  \label{fig:effectivearea}  
\end{figure}
\begin{figure}[H]
\centering
  \includegraphics[width=0.48\textwidth]{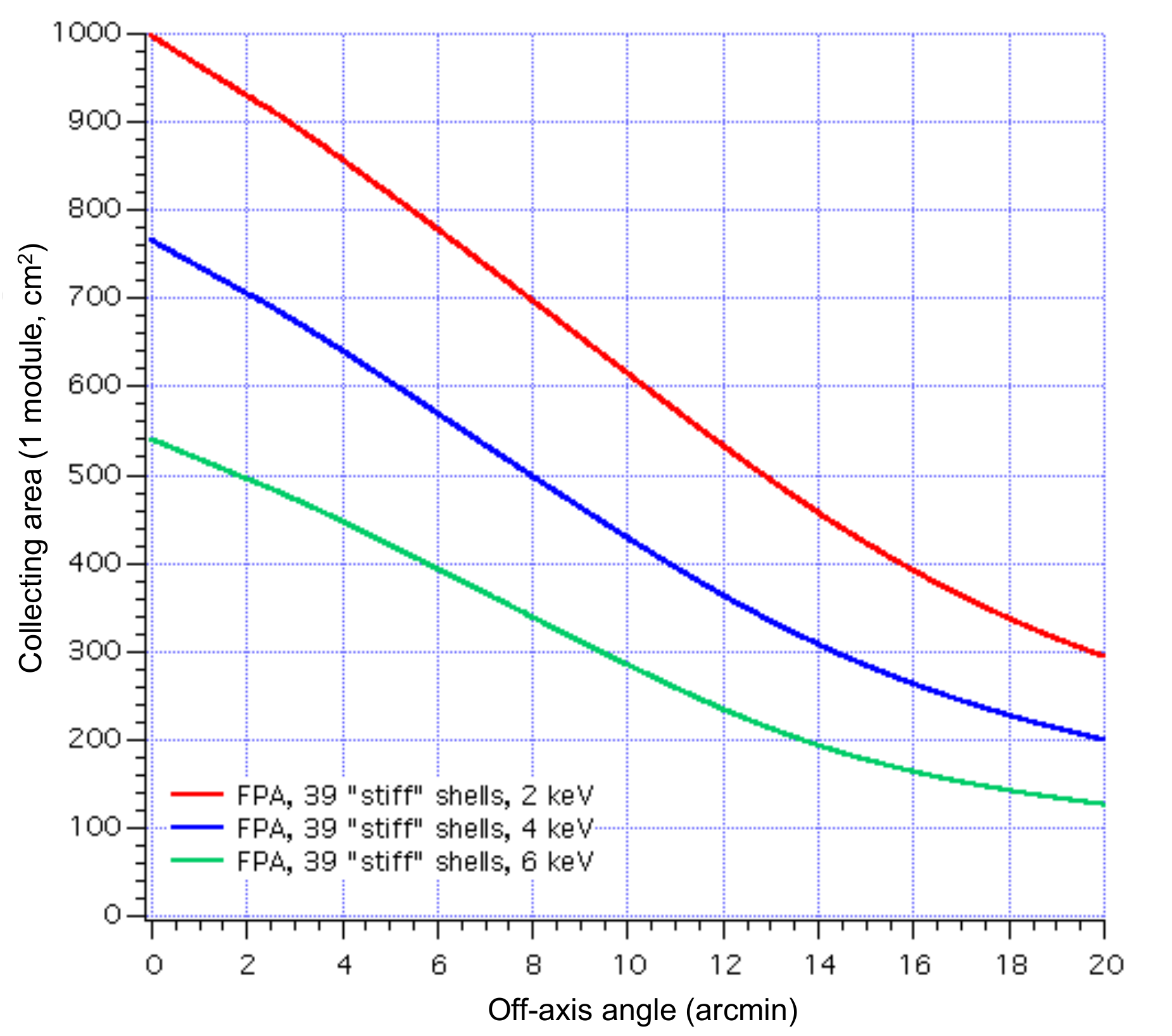}
\caption{Vignetting function at different X-ray energies or the eXTP X-ray baseline optics based on Nickel electroforming. The case of 39 stiff shells is presented.}
  \label{fig:vignetting}  
\end{figure}
A set of 39 mandrels at 60 cm length, parabola + hyperbola (30 cm x 30 cm) monolithic pseudo--cylindrical mandrels will be developed for the four shells of PFA and all SFA modules. The diameters of the shells range from 507 to 267 mm. To meet the different requirements for the angular resolution, and to optimize the mass constrains, two different thickness-to-diameter ratio sequences for the Nickel mirror walls of the shells are considered \cite{Basso1,Basso2}.

For the PFA optics, the thickness--to--radius is expected to be 0.00175, with a minimum and maximum thickness of 0.231~mm and 0.438~mm respectively. The weight of each module is about 82~kg (for the 39 shells) + 24~kg (for the structure), so that the total mass of the 4 modules is approximately 416~kg. For the SFA optics, the thickness--to--radius is about 0.0009, with a minimum and maximum thickness of 0.120~mm  and  0.225~mm respectively. The weight of each module is about 43~kg for the 39 shells and 13~kg for the structure, for a total of 504 kg for 9 modules. 

The expected collecting area for a single X-ray optics module (both polarimetric and spectroscopic) is presented in Fig.~\ref{fig:effectivearea}, while in Fig.~\ref{fig:vignetting} the vignetting function, inferred using well-tested simulation software tools \cite{Spiga1,Spiga2}, is shown. Preliminary analyses have been performed on the thermo-structural behaviour. Very promising results in terms of resistance to static stress and natural frequencies excitations have been obtained using solutions based on single or double spiders.

\textit{Slumping Glass Optics} Slumping glass optics (SGO) is the second technology considered for the eXTP optics. The design is based on the Wolter-I concept. Each mirror is fabricated by a thermal slumping process and coated with reflective thin films. The mirror length is 100~mm. To achieve the required collecting area around 200 shells need to be assembled with diameters from 110 to 500~mm. The number of shells can be significantly reduced by developing longer glass mirrors. An important advantage of the SGO technique is the light weight of the mirrors since their density is only $\sim$2.5 g/cm$^{3}$. 
Thermally slumped glass optics have been used for the NuSTAR mission, whose telescopes contain 133 shells of thin glasses \cite{Hailey2010, Chan2014}. During the slumping process, the glass is slumped very slowly onto the convex surface of the mandrel to assume its shape \cite{Zhang2009}. The temperature curve needs to be optimized to control the slumping process and the figure of the glass. Two furnaces temperature control of less than $\pm$1 $^{\circ}$C and excellent uniformity have been used for the preliminary study and production of cylindrical glass substrates. Schott D263, 200-300 $\mu$m thick, is the material used as slumping glass. To fabricate glass mirrors with high accuracy, even higher accuracy metrology tools are required. Both the mandrel and the glass are measured by a laser scanning system and an interferometer. 

In the context of the XTP study, cylindrical glasses with 50-60 arcsec resolution (HPD) have been produced. The best mirror produced has an angular resolution of 36 arcsec (HPD). 



\begin{figure}[H]
\centering
\includegraphics[width=0.8\columnwidth]{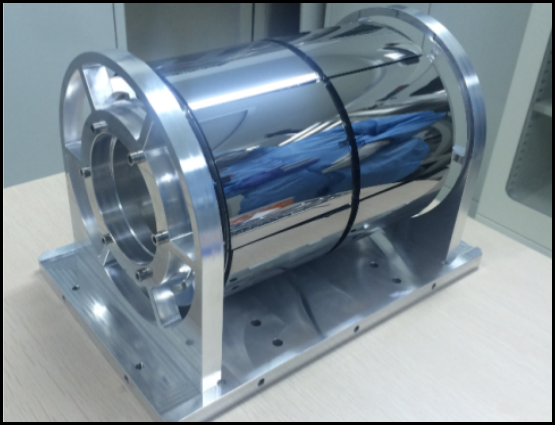}
\caption{The first prototype of the telescope based on slumping glass optics.}
\label{Figure:SGO_prototype}
\end{figure}

To increase the reflectivity in the range 0.5-10~keV the slumped glass substrates are coated with Platinum and Carbon layers. The thicknesses of the layers are optimized for each shell to maximize the collecting area. Measurements performed with the AFM and optical profiler have resulted in a surface roughness of 0.32-0.37~nm. The X-ray reflectance has been measured to be of 90\% (70\%) at 1 (8) keV.  
The SFA module is built from individual glass segments. To achieve large collecting surfaces, concentric shells of glasses are stacked on top of each other starting from the central mandrel. In this process, the cylindrical glass segments are forced to a conical form while the middle frequency figure and microroughness of the glass are maintained. The  advantage of this process is that each glass shell is machined with respect to the optical axis and not the last shell so there is no significant stack-up error due to assembly. The assembly system integrates both single point diamond tools and abrasive (grinding) wheels, which allows to reach sub-micron accuracy. Each segmented mirror is bonded to the structure using graphite spacers and epoxy with a fine optimized air pressed instrument. Metrology plays an essential role during the assembly. We have developed three techniques to measure the geometry parameters of the module, including contact mechanical probe and non-contact optical probe. The different probes combined with the machine’s ultra-precision stage calibrate each other to ensure an optimal measurement accuracy. Based on the metrology results of the free-standing mirrors and after assembly, the impact of assembly error on the angular resolution is controlled within 1 arcmin.  Prototypes of the SFA optics have been developed by the Tongji University, China. The 1st complete shell prototype has a focal length of 4 meters and 5 shells, with the outer shell diameter of 170 mm as shown in Figure \ref{Figure:SGO_prototype}. Sixty shells coated with a Platinum have been assembled. The prototype has been tested at the National Astronomical Observatory of China. The HPD of the spot has been estimated to be 3 arcmin at 2-8 keV. The angular resolution of a complete shell prototype is estimated to be $\sim$1.5~arcmin.

\subsubsection{SFA Detectors and Electronics}
\begin{figure}[H]
\centering
\includegraphics[width=0.8\columnwidth]{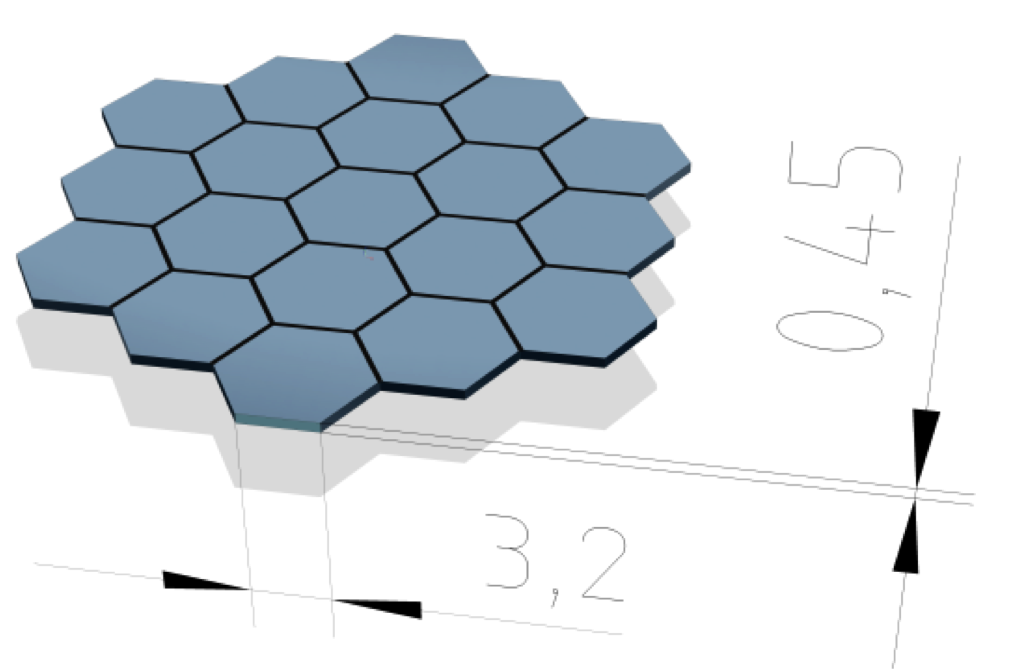}
\caption{SDD array for the SFA detectors. Each of the 19 hexagon cells has a side length of 3.2\,mm and a sensitive thickness of 450\,$\mu$m.}
\label{Fig:SFA_elec}
\end{figure}

\textit{The SFA Detector.} The scientific objectives of the SFA require a detector with high count rate capability and time resolution, along with excellent spectroscopic performance. The energy resolution is required to be better than 180 eV (FWHM at 6 keV) until the end of the mission after 8 years lifetime, and time resolution shall be less than 10\,$\mu$s. The baseline detector consits of an array of SDDs organized in 19 hexagonal cells. This configuration meets the aforementioned requirements (see Fig.~\ref{Fig:SFA_elec}). Compared with CCDs, SDDs allow a much faster readout. The detector geometry ensures that the vast majority of source photons are focused onto the inner 7 cells of the array, whereas the cells of the outermost ring are used to accurately determine the background. The side length of a hexagon cell has thus been determined to be 3.2~mm matching the expected angular resolution of the optics. The detector has a 450\,$\mu$m sensitive thickness, which delivers excellent quantum efficiency over the SFA energy range. Blocking of optical photons is achieved by coating the silicon sensor entrance window  with a thin Aluminum layer of $\sim80$\,nm thick.  The SDD sensors will be provided by the Max-Planck-Institut für Extraterrestrische Physik, Garching (Germany) in collaboration with the Semiconductor Laboratory of the Max Planck Society. The fabrication of prototype SDD arrays is currently ongoing. Each SDD pixel has a detecting area of 26.6\,mm$^2$, and the whole detecting area of the SDD array is 5\,cm$^2$. The depletion layer is 450\,$\mu$m. A metal mask is planned to be implemented above the detector, to cover the gaps, thus decreasing the ratio of split events generated at the edge. A passive shield is used to reduce the background. This consists of a deflector assembled under the mirror module, a baffle installed above the filter wheel, and multi-layer composites glued around the detector with the exclusion of the entrance window.

\textit{Front End Electronics.} The signal processing chain of each SDD cell includes a charge sensitive pre-amplifier, a fast shaper, and a slow shaper with sample and hold circuit. Since most of the incident photons will be focused onto the central cell of the SDD array, a dedicated analogue to digital conversion (ADC) circuit is implemented to reduce dead time effects. For the remaining 18 cells, to simplify the readout, a multiplexed solution is considered, i.e., every 6 cells share one ADC circuit. The fast shaping time constant is $\sim0.2$\,$\mu$s time to allow an accurate event time determination and pile-up rejection. A slow shaper with a time constant of several $\mu$s allows precise signal charge measurement and therefore a better spectral resolution. The concept hass been demonstrated with discrete components already, and a more compact solution based on dedicated ASICs is currently being developed.
\begin{figure}[H]
\centering
\includegraphics[width=0.8\columnwidth]{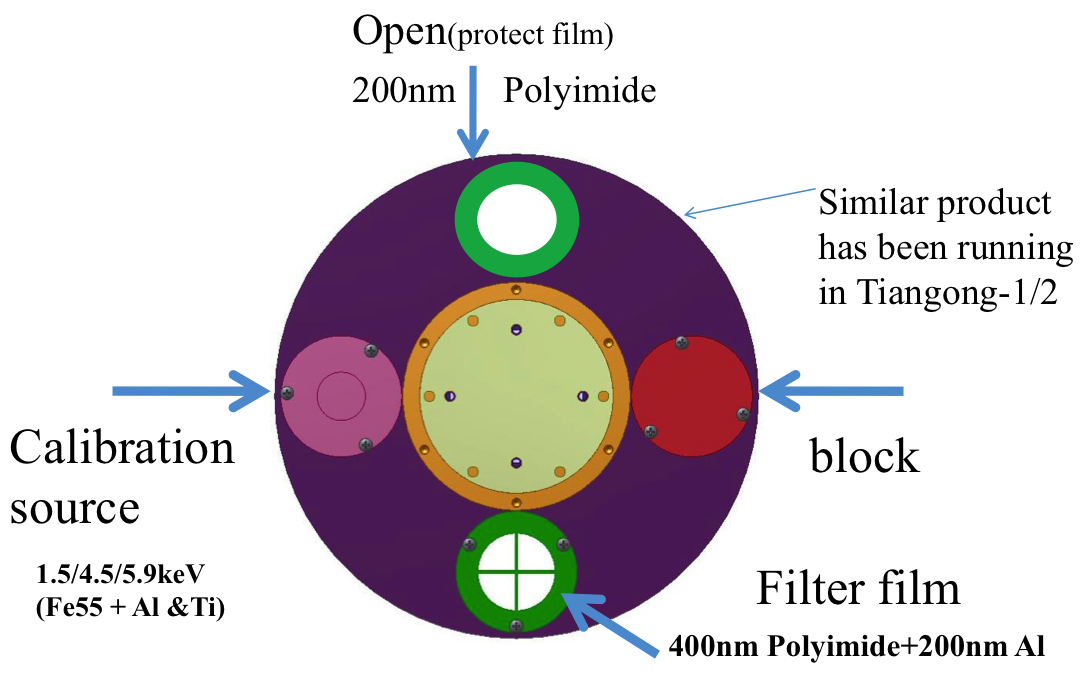}
\caption{The schematics of the filter wheel.}
\label{fig_sfa_wheel}
\end{figure}
 
\textit{Back End Electronics.} An FPGA-based back-end electronics (BEE) module is implemented to control the front-end electronics (FEE), and readout the digitized data from the ADCs. For each incident X-ray photon, the signal amplitude and arrival time are recorded and sent to the spacecraft data handling subsystem for temporary storage and telemetry. The FPGA also generate housekeeping data packages that include key parameters like e.g., counting rate of each cell, live time of the readout circuit, and working temperatures.

\subsubsection{The Filter Wheel and Cooling}
A filter wheel is located above the detector (see Fig.~\ref{fig_sfa_wheel}). In the current configuration, 4 positions corresponding to 4 operational modes are foreseen: (1) \textit{Open filter}, used in ground tests and weak sources observations; (2) \textit{Calibration}. An Fe-55 radioactive source is used for in orbit calibration; (3) \textit{Optical blocking filter}. This filter, consisting of a 400\,nm polyimide and a 200\,nm Aluminum film, is used to block visible and UV light. It also prevents contamination of the detector; (4) \textit{Closed filter}. A metal shutter is used to measure the internal background, and prevents the detector from damage during launch and extreme solar events. To compensate for particle damage of the SDDs, especially due to protons, a low temperature from -100 to -80\,$^{\circ}$C is needed during the operation of the detectors. Temperature stability of $\pm$0.5\,$^{\circ}$C is also required. An active method is designed to achieve the low temperature: it includes a Helium pulse tube and loop heat pipes.
\subsubsection{Performance: Effective Area and Background}
\begin{figure}[H]
\centering
\includegraphics[width=1.\columnwidth]{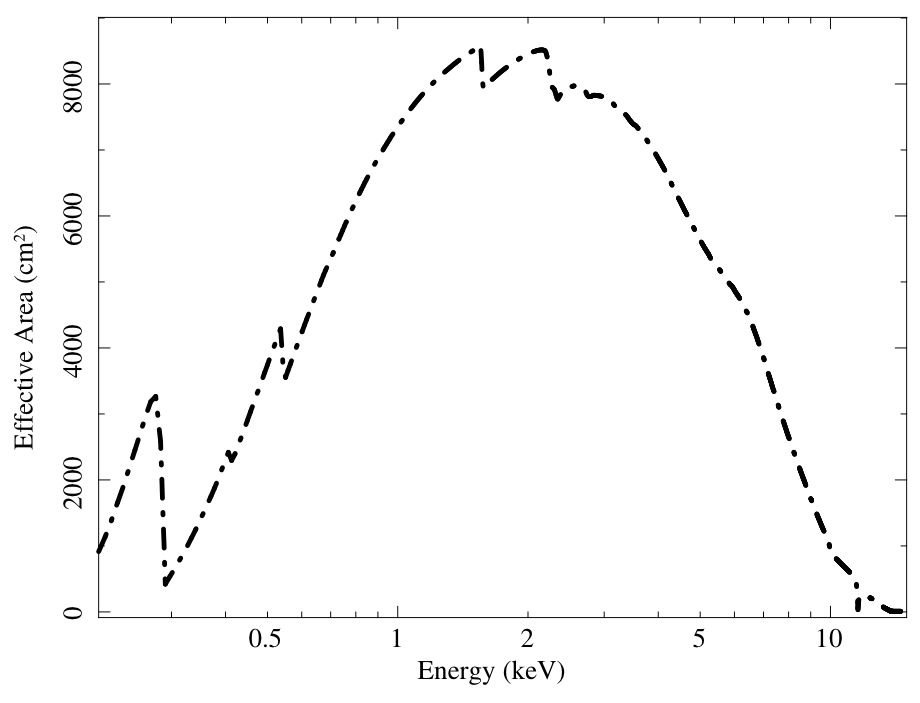}
\caption{The effective area of the SFA. The efficiency of the detectors and the filters is taken into account.}
\label{fig_sfa_eff_area}
\end{figure}

\begin{figure}[H]
\centering
\includegraphics[width=1.\columnwidth]{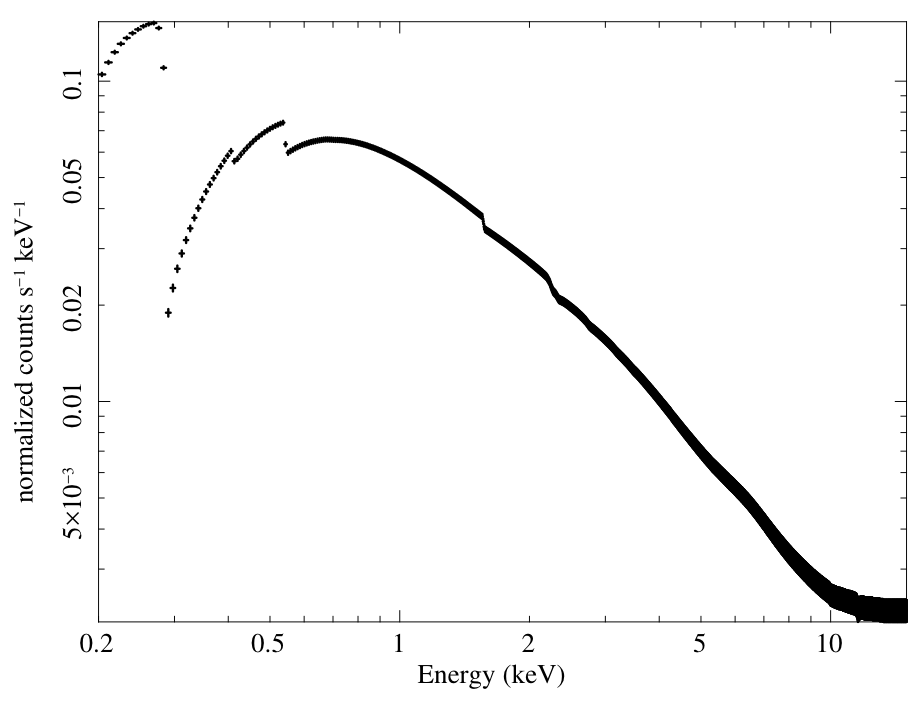}
\caption{Expected background of the SFA.}
\label{fig_sfa_bck}
\end{figure}

The total effective area of the SFA, taking into account the detector efficiency and filters, is shown in Fig.~\ref{fig_sfa_eff_area}. Similarly, Fig.~\ref{fig_sfa_bck} shows the background expected on the basis of current simulations. 

\subsection{Large Area Detector - LAD}
 \label{sec:LAD}
The LAD is designed to perform, on a large collecting area, photon-by-photon observations of X-ray sources in the energy range of 2-30 keV (up to 80 keV in the expanded mode, for out-of-field-of-view burst events). The arrival time and energy of each photon are measured with high resolution and for a very large statistics of events. This enables unprecedented spectral-timing studies in the X-rays. The FoV is limited to $\sim$1$^{\circ}$ by a mechanical collimator to reduce source confusion and the X-ray background. The key innovation of the LAD (as well as of the WFM, see section \ref{sec:WFM}) relies on the technology of large-area SDDs and capillary plate collimators. These elements allow the construction of highly efficient X-ray detectors, which are only a few mm thick and a few hundred grams in weight, thus reducing requirements for weight, volume and power by about an order of magnitude with respect to proportional counters and mechanical collimators used in the past generation of large area instruments \cite{Jahoda1996}. The LAD main specifications are listed in Table \ref{tab:LAD}.

\begin{table}[H]
\begin{center}
\caption{LAD main specifications.}
\label{tab:LAD}
\footnotesize
\begin{tabular}{l|l}
\bottomrule
\textbf{Parameter} & \textbf{Value} \\
\hline
Energy Range (nominal) & 2–30 keV (extended 30-80 keV)\\ 
Effective area & 3.4 m$^{2}$~at 8~keV \\
 & 0.37 m$^{2}$~at 30~keV\\
Energy resolution (FWHM, at 6 keV) & $<260$~eV (all events)\\
 & $<200$~eV (40\% of events)\\
Field of View & $<60$~arcmin \\
Time resolution& 10~$\mu$s\\
Dead Time (at 1 Crab) & $< 0.5$\%, (goal $< 0.1$\%) \\
Background & $< 10$~mCrab\\
Maximum source flux (continuous) & $>300$~mCrab\\
Maximum source flux (for 8 hrs) &$>15$~Crab \\
Total mass & $\sim360$~kg\\
Power& $\sim400$~W\\
\bottomrule
\end{tabular}
\end{center}
\end{table}

\subsubsection{The LAD architecture}
\label{sec:LAD_architecture}
\begin{figure}[H]
\includegraphics[width=1.0\columnwidth]{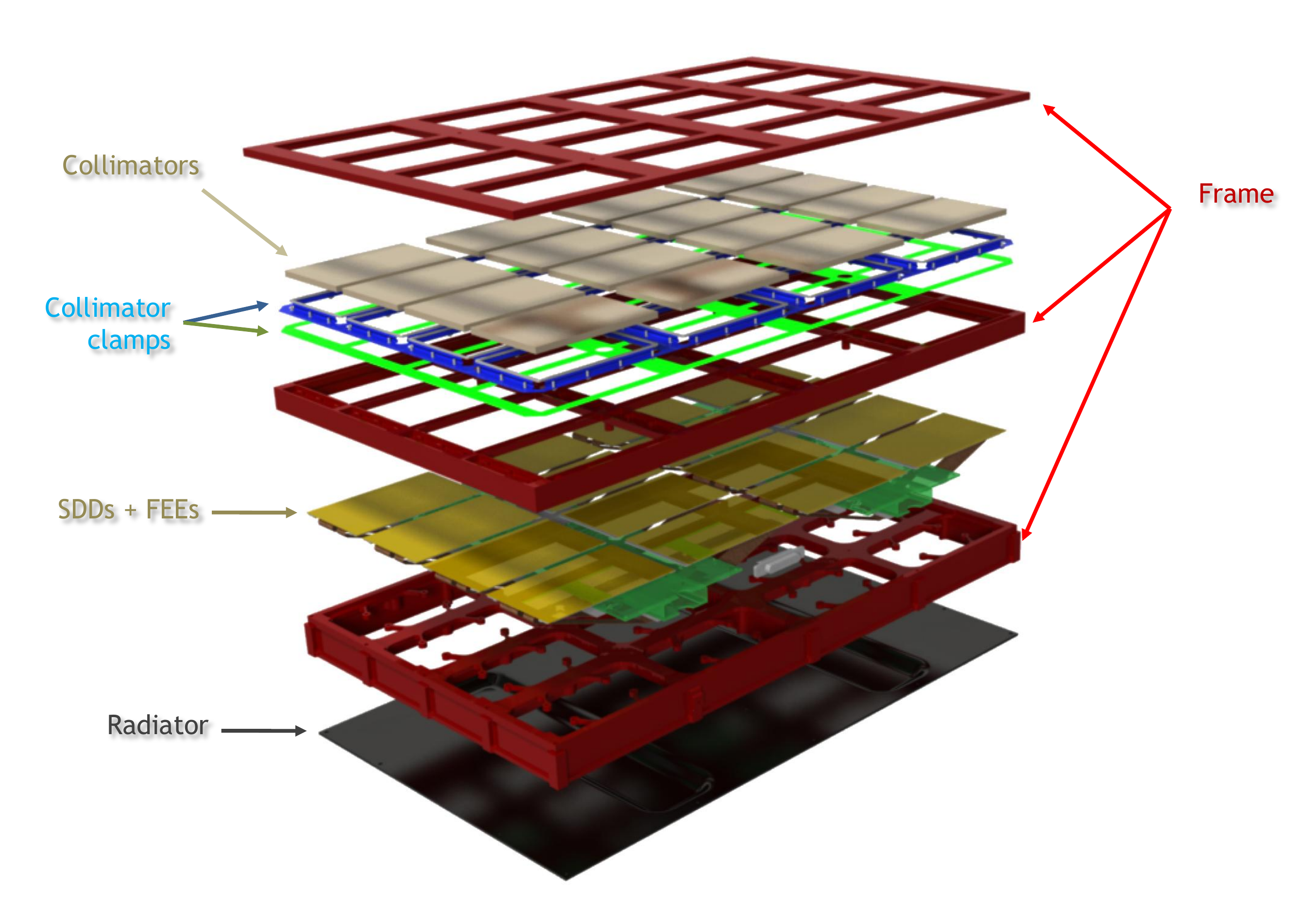}
\caption{LAD module exploded.}
\label{fig:module_exploded}
\end{figure}

The LAD large area is achieved by a modular and intrinsically highly redundant design. The fundamental units of the instrument are the LAD modules. Each module consists of a set of 4~$\times$~4 large area SDDs and 4~$\times$~4 capillary plate collimators, supported by two grid-like frames. Each of the SDDs are equipped with the FEE, and interfaced to a support frame, the detector tray. The detector tray hosts the module back-end electronics (MBEE) on its back-side, which includes the digital electronics and power supply units (PSUs). The detector tray and the collimator tray are bolted together to form the complete LAD module. The module and its main components are shown in Figure \ref{fig:module_exploded}. Modules are embodied in two panels, Carbon-fiber frames deployed from the optical bench. Each of the panels hosts 20 (5~$\times$~4) modules, for a total of 40 modules or 640 detectors. Part of the panel is the panel back-end electronics (PBEE), that interfaces the 20 Modules to the central instrument control unit (ICU).

To summarize, the LAD module consists of: 
\begin{itemize}
\item A collimator tray, containing 16 co-aligned collimator plate tiles (one per SDD, see Section \ref{sec:collimators}). 

\item The detector tray, containing 16 SDDs and the FEEs. It interfaces the MBEE through a built-in rigid-flex harness and hyperstac connectors.

\item The MBEE, organized in two sections, each interfacing 8 detectors. It is located on the back side of the detector tray. It controls the SDDs, FEEs and PSUs, reads out the FEE digitized events, generates housekeeping data and ratemeters, formats and time-stamps each event, and transmits it to the PBEE.

\item A PSU located within the same structure of the MBEE. It converts the supply power to low (3V), medium (100V) and high (1.3kV) voltages for the SDDs, FEEs and ASICs.

\item A 300 $\mu$m Lead back-shield, to reduce the background events in the SDDs, and a 2 mm Aluminum radiator to dissipate heat from the module to the backside of the module box.

\end{itemize} 

The LAD panel structure in Carbon Fiber Reinforced Plastic (CFRP) supports the LAD modules and the PBEE. The key requirements for this structure are to withstand the launch, provide mechanical and thermo-elastic stability to the modules (e.g., alignment), and aid the thermal control. To this purpose a common radiator is used on the back side of the panel. The interfaces to the module, optical bench and mechanisms are in Titanium. A sun shield protects the LAD panels from direct solar irradiation. Twenty modules are controlled by two PBEEs, located underneath the panel near the hinge to minimise the amount of harness that has to cross the hingeline. The PBEE sends commands to the MBEEs and distributes clocks and power. It receives data from the MBEEs and sends them to the ICU (see Section \ref{sec:lad_detectors_and_electronics}).  The ICU is cold redundant. In total, the LAD is composed of 640 SDDs, 640 CPs and 5120 ASICs (8 per SDD). To monitor the internal background, one out of 40 modules is equipped with a ``blocked collimator'' made of the same materials of the other collimators but with no pores. 

\subsubsection{Detectors and electronics} 
\label{sec:lad_detectors_and_electronics}

\begin{figure}[H]
\includegraphics[width=\columnwidth]{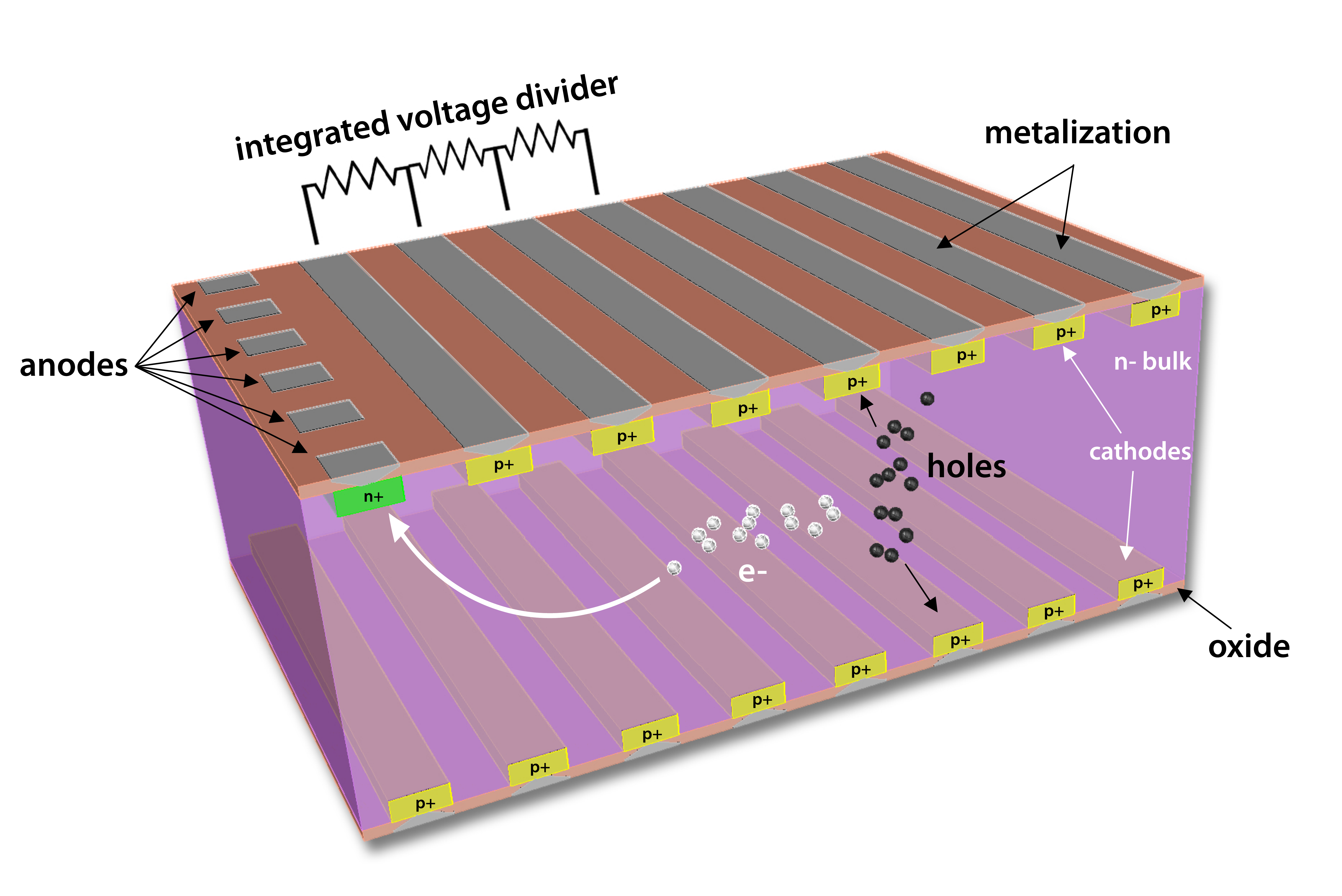}
\caption{SDD working principle.}
\label{fig:sdd_working_principle}
\end{figure}

\textit{The SDDs}. Large-area SDDs \cite{gatti84} were developed for the inner tracking system of the ALICE experiment at the large hadron collider \cite{vacchi91}. They were later optimized for the use as photon detectors onboard LOFT \cite{rachevski14} with typical size of 11~$\times$~7 cm$^{2}$ and 450~$\mu$m thickness. SDDs are capable to read-out large photon collecting areas with a small set of low-capacitance (and thus low-noise) anodes, and are light with $\sim$1 kg m$^{-2}$. The working principle is shown in Figure~\ref{fig:sdd_working_principle}. The cloud of electrons produced by the interaction of the incident photon drifts towards the read-out anodes at $\sim$0~V, driven by a constant electric field generated by a progressively decreasing negative voltage applied to the cathodes. Diffusion causes the electron cloud to expand by a factor depending on the square root of the drift time. The charge distribution over the collecting anodes thus depends on the absorption point in the detector.

While drifting and diffusing, the electron cloud size increases and, at a distance $d$ from the photon absorption point, the typical size of the Gaussian-shaped cloud is given by $\sigma \simeq \sqrt{ \frac{2k_bT}{qE} \times d}$, where $k_{b}$ is the Boltzmann constant, $T$ is the temperature, $q$ is the electron charge, and $E$ is the drift electric field. The cloud arriving at the anodes can be described by a Gaussian function, with an area $A$ equal to the total charge (i.e the photon energy), a mean value $m$ representing the "anodic" coordinate of the impact point, and a widht $\sigma$ that depends on the absorption point. 

Each LAD detector is segmented in two halves, with 2 series of 112 read-out anodes (with~970~$\mu$m pitch) at the two edges and the highest voltage along its symmetry axis. A drift field of 370~V/cm (1300~V maximum voltage) gives a drift velocity of $\sim$5~mm/$\mu$s and a maximum drift time of $\sim$7~$\mu$s. This is the largest contribution to the uncertainty in the measurement of the arrival time of the photon. 
The segmentation into 640 detectors and $144\times 10^{3}$ electronics channels insures that the rate on the individual channel is very low even for very bright sources. Pile-up or dead-time effects are therefore negligible. On an equatorial orbit, to maintain the required energy resolution until the end of life, the detectors need to be moderately cooled (-10 $^{\circ}$C) to reduce the leakage current. Considering the large size of the LAD this is achieved with passive cooling. 

\begin{figure}[H]
\includegraphics[width=1.0\columnwidth]{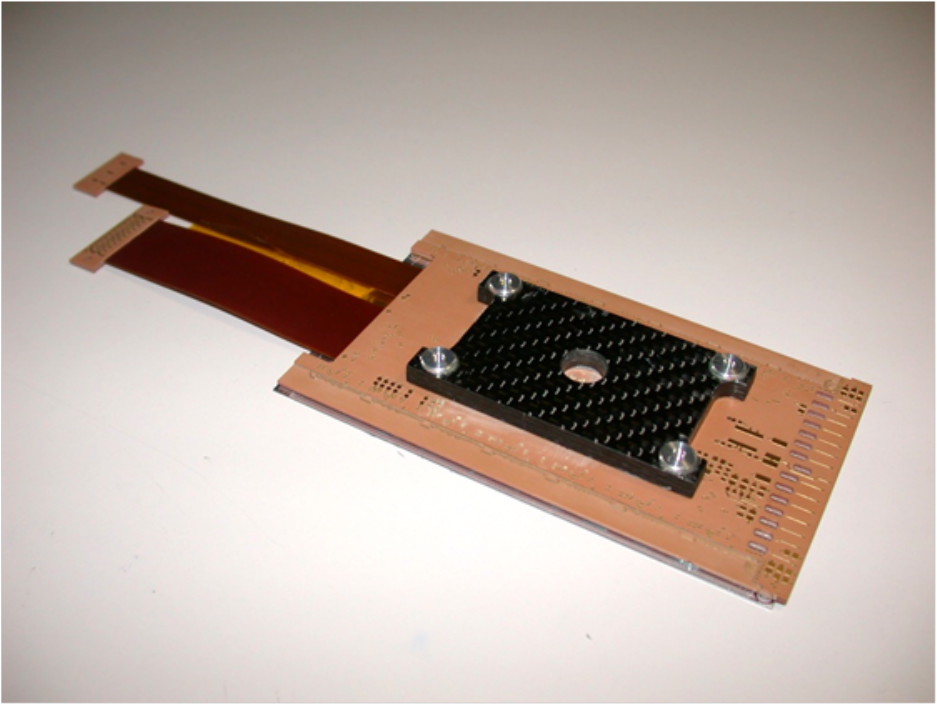}
\caption{The protytpe of the LAD FEE realized for the LOFT study.}
\label{fig:FEE_prototype}
\end{figure}

\textit{The} ASICS. For the high-density read-out of the detector dedicated ASICs with excellent performance and low power (requirement is 17 e$^{-}$ rms noise with 650 $\mu$W per channel) are needed. The read-out is performed by full-custom 8$\times$32-channel IDeF-X ASICs, inherited from the IDeF-X HD and IDeF-X BD ASICs successfully used in the ESA's Solar Orbiter Mission \cite{Gevin2012}, \cite{Gevin2016} with A/D conversion carried out by one 16-channel OWB-1 ASIC for every detector \cite{Bouyjou2017}. The dynamic range of the read-out electronics is required to record events with energy up more than 80 keV. Events in the nominal energy range (2-30 keV) are transmitted with 60 eV binning, while those in the extended energy range (30-8 keV) are transmitted with reduced energy information (2 keV wide bins) as they will be used to study the timing properties of bright/hard events shining from outside the FoV (e.g., gamma-ray bursts, magnetar flares). 
\begin{figure}[H]
\includegraphics[width=1.0\columnwidth]{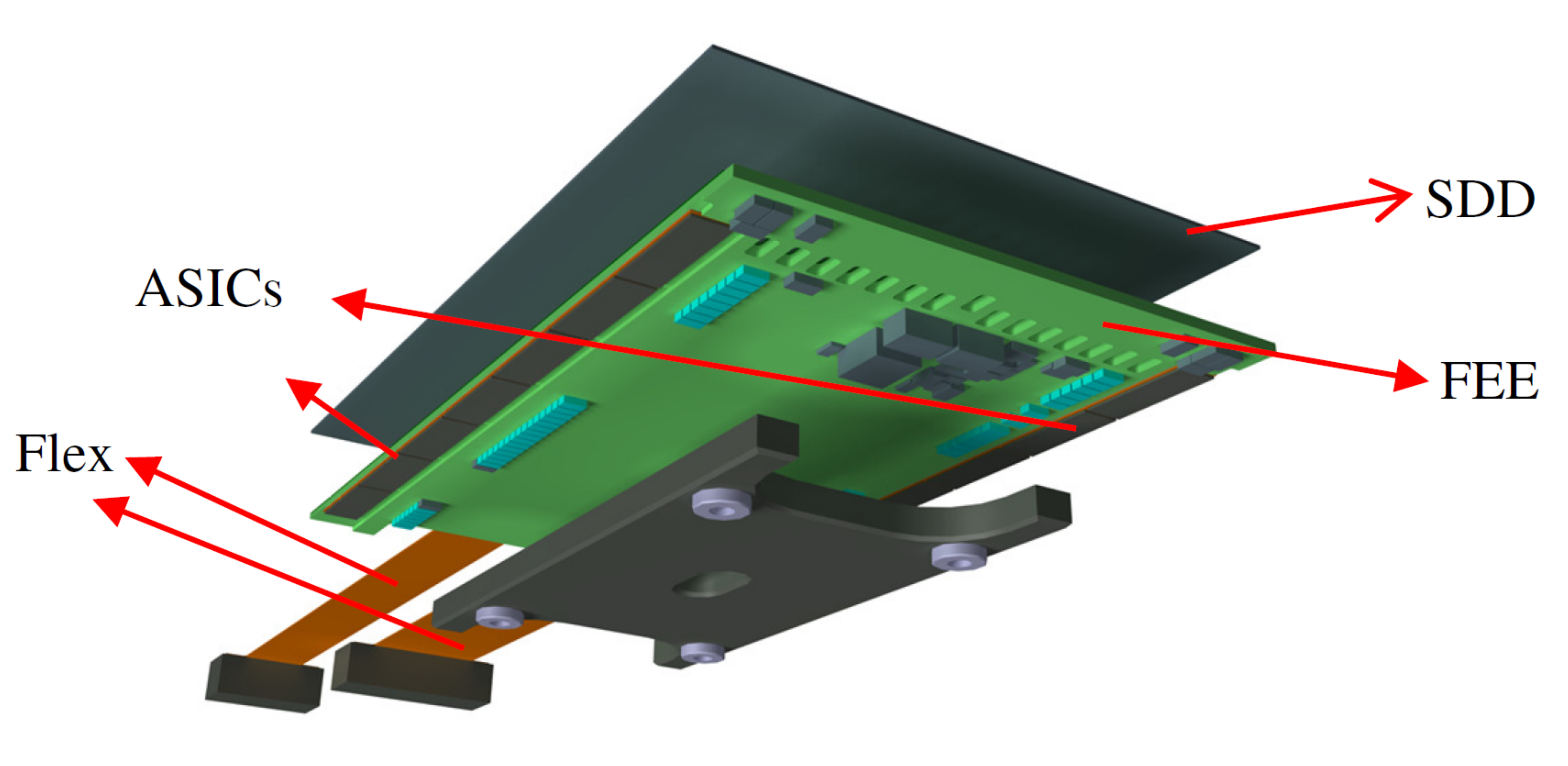}
\caption{A schematic of the LAD Front End Electronics.}
\label{fig:FEE_exploded}
\end{figure}
The read-out ASICs are integrated on a rigid-flex PCB forming the FEE, with the task of providing filtered biases to SDDs and ASICs, I/O interfaces, and mechanical support and interface to the module. The SDD will be back-illuminated, allowing for direct wire-bonding of the anode pads to the ASIC input pads. The flat cable connection to the MBEE is part of the rigid-flex PCB structure. A view of the CAD design of the LAD FEE and the mechanical prototype realized for the ESA's LOFT study are shown in Figures \ref{fig:FEE_exploded} and \ref{fig:FEE_prototype}.

\textit{Back-End Electronics.} When the collected signal charge exceeds a programmable threshold in one of
the ASIC channels, a trigger is generated and forwarded to the BEE. In case of a
confirmed valid trigger pattern, the collected signals in all ASICs of the respective detector half are then
digitised and passed on to the BEEs. Following the A/D conversion, the BEE event processing pipeline is activated: a time tag is added to each event, and a pedestal and common noise subtraction is performed. In addition,
energy reconstruction takes place to determine the event parameters. Besides the event processing, the
BEE controls the FEE and PSU operation, generates housekeeping data and rate meters, and transmits
formatted event packets onward. The LAD BBE is organised into two hierarchical levels, due to the number of
detectors in the design. The MBEE will be located on the back
side of the detector tray and is organised in two PCBs, interfacing 8 detectors each. The central
component of each MBEE PCB is an RTAX-SL FPGA. The PBEE located on the back side of the panel interfaces each of the twenty
MBEEs and the ICU. The two PBEEs also make use of the same FPGA and
collect and buffer the received event packets, reorganise the data depending on the observation mode,
and transfer the data to the data handling unit (DHU) of the ICU along with the housekeeping data via a
SpaceWire interface.

\textit{Instrument Control Unit.} The ICU forms the central controlling element of the instrument. It provides the
interface to the spacecraft OBDH and also access to all instrument sub-systems via SpaceWire. The ICU
box consists of three components: the DHU, the mass memory, the power distribution unit (PDU). 
Standard tasks performed at the ICU level involve telecommand execution and distribution,
access to mass memory, time distribution and synchronisation, data processing and compression, housekeeping data
collection, instrument health monitoring and calibration tasks. The ICU box contains each PCB board twice for cold redundancy.

\subsubsection{Collimators, Filters} 
\label{sec:collimators}
\begin{figure}[H]
\includegraphics[width=1.0\columnwidth]{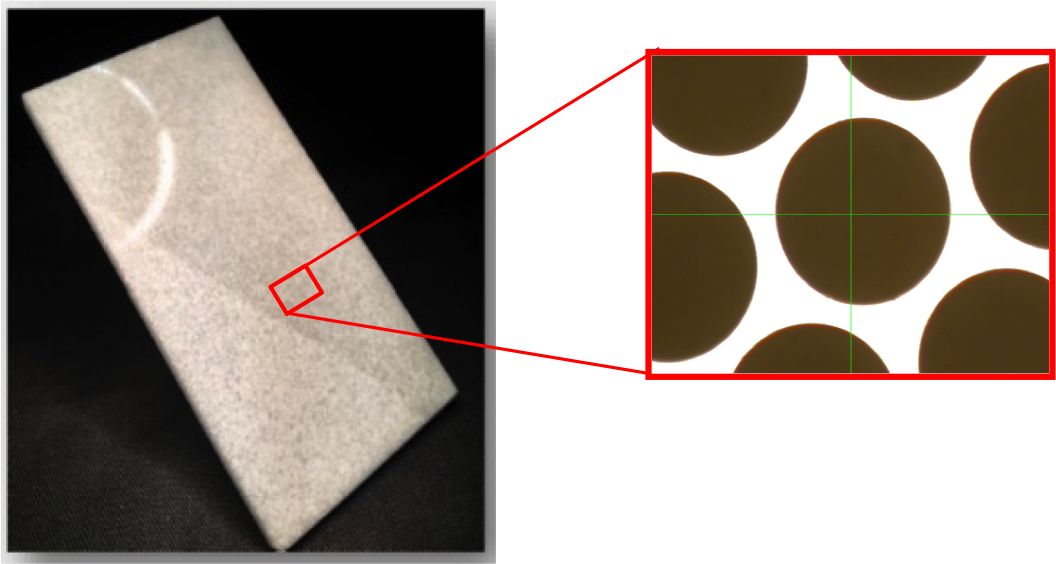}
\caption{A prototype of the LAD Collimator developed by NNVT.}
\label{fig:LAD_Collimator}
\end{figure}

To get full advantage of the compact detector design, a similarly compact collimator design is provided by the mechanical structure of the mature technology of the micro-channel plates, the capillary plates.
In the LAD geometry, it is a 5~mm thick sheet of lead-glass ($>$40$\%$ Pb mass fraction) with the same dimensions of the SDD detector, perforated by thousands of round micro-pores with a 83 $\mu$m diameter, limiting
the FoV to 0.95$^{\circ}$ (FWHM). The open area ratio of the device is $\geq$75$\%$ (Figure \ref{fig:LAD_Collimator}).
The thermal and optical design is then completed by an additional optical filter, composed by a thin (1 $\mu$m thickness) Kapton foil coated on both sides with 40 nm of Aluminum. This is to guarantee a 10$^{-6}$ filtering on IR/Visible/UV light, while transmitting $>$90$\%$ of the light at 2 keV and above.

\subsubsection{Performance: Area and Background}
eXTP will be launched into an almost equatorial low-earth orbit at an altitude of $\sim$550~km. At these latitudes, the geomagnetic field effectively screens primary cosmic rays up to energies of a few GeVs. Moreover, the satellite crosses the South Atlantic Anomaly only in its external regions, thus minimizing the activation of materials. The main sources of background considered in the simulation studies \cite{campana13} are: 1) \textit{cosmic diffuse X-ray background} \cite{gruber99}; 2) \textit{earth albedo} $\gamma$-rays. We assumed the albedo spectrum as measured by BAT \cite{ajello08}; 3) \textit{residual primary cosmic rays}, estimated from AMS measurements \cite{alcaraz00a, alcaraz00b}; 4) \textit{secondary cosmic rays}, estimated from AMS measurements \cite{alcaraz00a, alcaraz00b}, and analytically modeled by Mizuno et al. \cite{mizuno04}; 5) \textit{earth albedo neutrons}, modelled with the QinetiQ Atmospheric Radiation Model (QARM) \cite{lei04}; 6) \textit{Natural radioactivity}. Since the Lead-glass collimators contain Potassium, radioactivity of $^{40}$K has to be considered. This component will be largely reduced by adopting the capillary plate technology of NNVT, which is almost Potassium-free.

\begin{figure}[H]
\centering
\includegraphics[width=1.0\columnwidth]{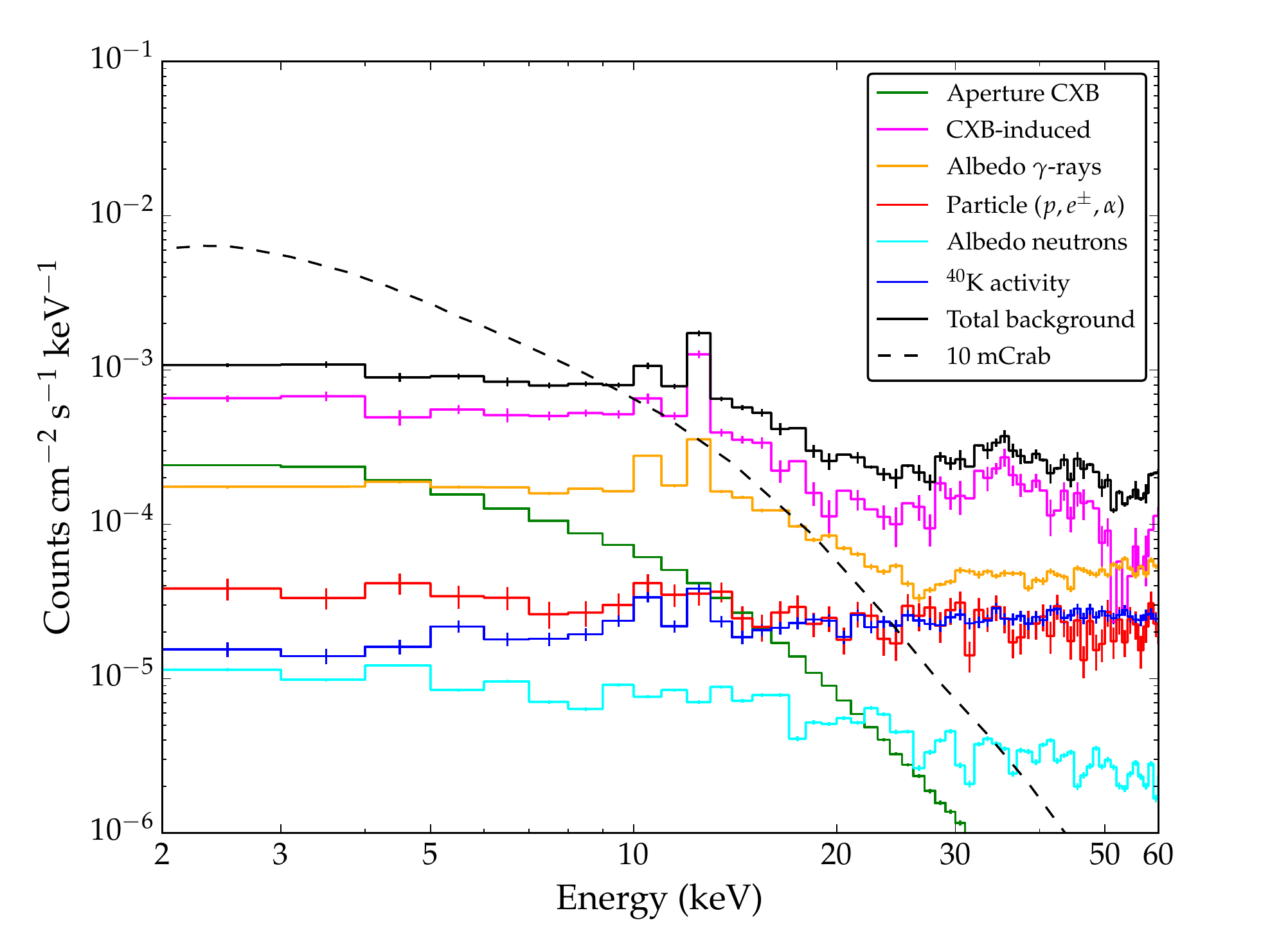}
\includegraphics[width=1.0\columnwidth]{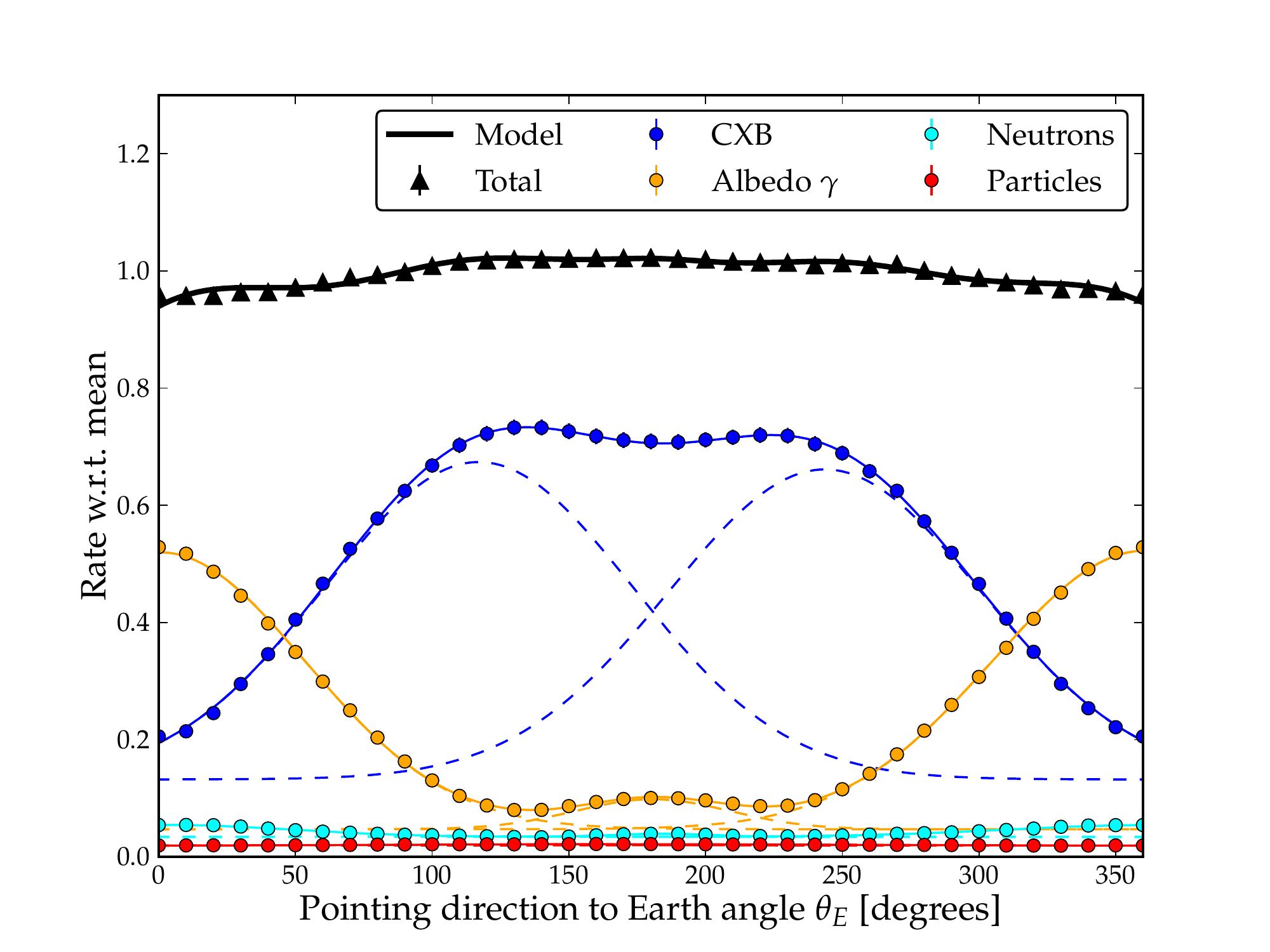}
\caption{Upper panel: Estimates of the LAD background. Lower panel: LAD background modulation. $\theta_E = 0^\circ$ corresponds to the earth's center aligned with the field of view, while $\theta_E = 180^\circ$ corresponds to the earth at the instrument nadir.}
\label{LADbkg}
\end{figure}
To estimate the LAD background we performed Monte Carlo simulations by using the Geant-4 Monte Carlo toolkit \cite{agostinelli03}. The total background is shown in the upper panel of Figure \ref{LADbkg}, together with  the spectrum of a 10~mCrab point-like source (dashed line). The main background contribution is due to CXB photons, that leak from and are scattered in the collimators or in the detector itself.  The diffuse emission collected through the FoV and the particle background are a minor contribution to the total count rate. Fluorescence emission from the Lead contained in the collimator glass ($L$-shell lines at 10.55 and 12.61 keV) and from the Copper contained in the FEE PCB board ($K$-shell lines at 8.05 and 8.90 keV) is present. The lower panel of Figure \ref{LADbkg} shows an evaluation of the background as a function of the angle between the LAD pointing direction and the center of the Earth. The maximum expected modulation of the background is $\sim$10\%. Since it is due to geometry, it can be predicted and modeled. By using a set of ``blocked'' detectors to monitor the instantaneous background, the overall background can be constrained to better than 0.5\%.
\begin{figure}[H]
\centering
\includegraphics[width=\columnwidth]{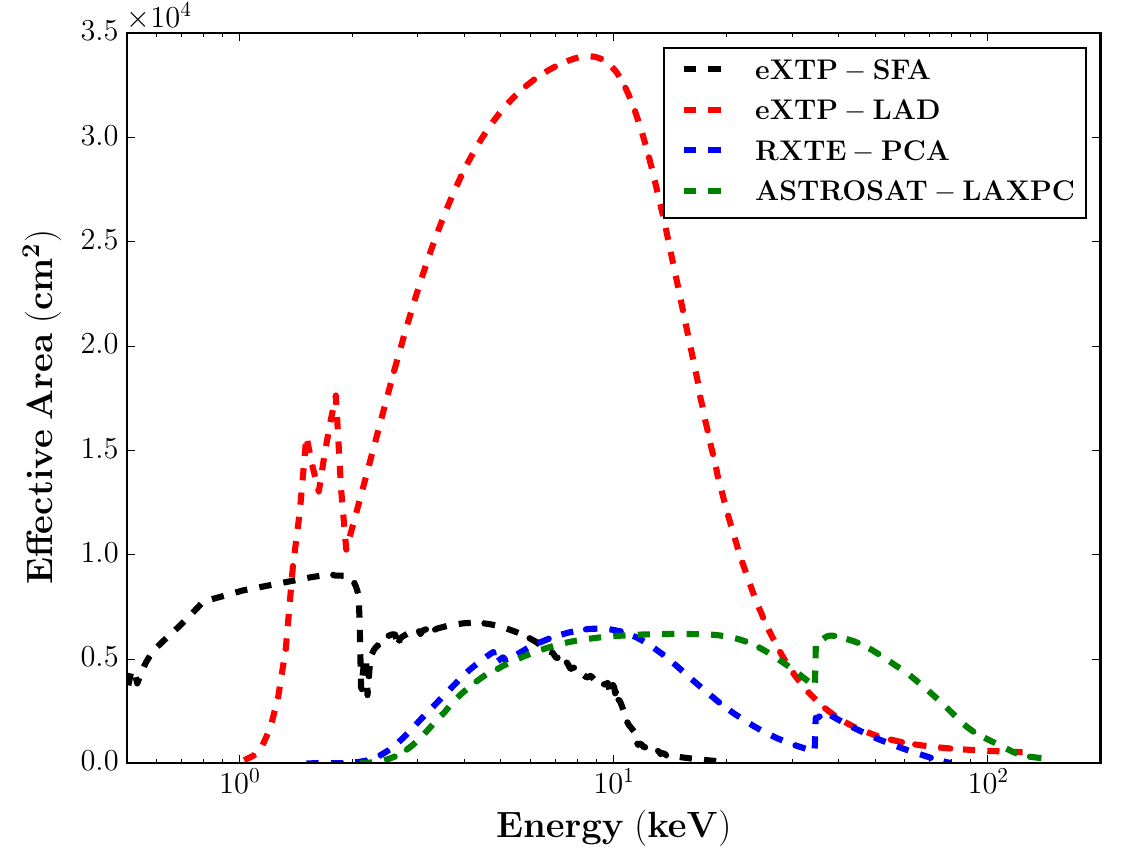}
\caption{The eXTP LAD effective area is shown, in comparison with that of the AstroSat-LAXPC and RXTE-PCA, the largest area instruments flown as of today. The eXTP-LAD area largely surpasses any past or currently flying mission and features an unprecedented value of about 3.4 m2 at 6 keV. In the figure the eXTP Soft Focusing Array effective area is also shown.  }
\label{LADeffective}
\end{figure}
\begin{figure}[H]
\centering
\includegraphics[width=0.9\columnwidth]{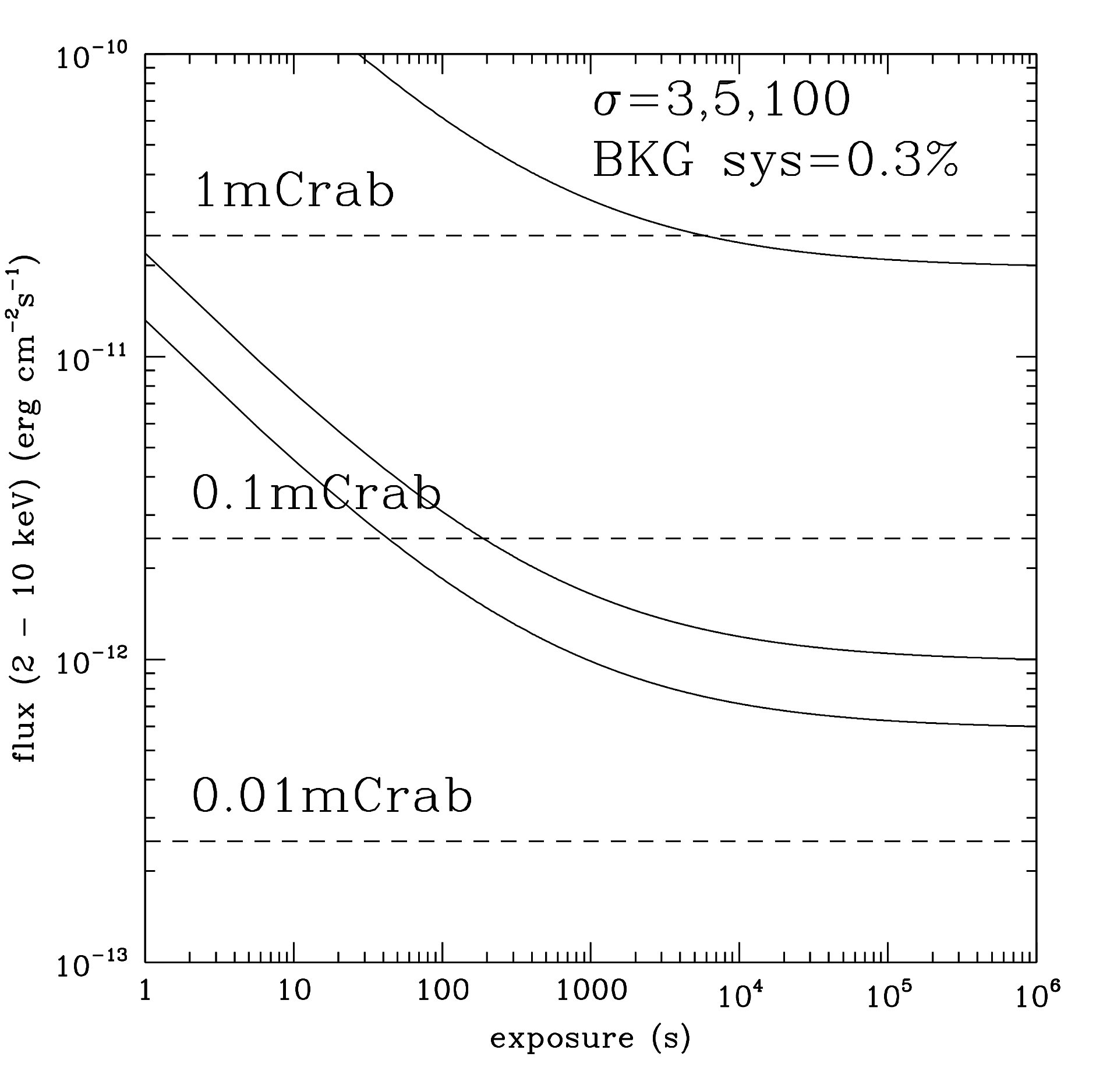}
\caption{The LAD minimum detectable flux (MDF) in 2--10 keV. The curves refer to a signal-to-noise ratio of 3, 5 and 100 respectively. We also included a 0.3\% systematics uncertainties on the background subtraction.}
\label{fig:lad_mdf}
\end{figure}

The effective area of the LAD, shown in Figure \ref{LADeffective} compared to that of the AstroSat-LAXPC and RXTE-PCA, the largest area instruments flown as of today, implies a count rate of $\sim$80,000 cts/s for a Crab-like spectrum. In Figure \ref{fig:lad_mdf} we show the minimum detectable flux (MDF) in the 2--10 keV range accounting for a 0.3\% systematics uncertainty on the background subtraction. These curves show that the statistical LAD 3$\sigma$ sensitivity for persistent sources is exceptionally high and amount to $\sim0.5$~mCrab/s. Due to the enormous amount of collected photons, for exposures larger than 10$^4$ s, the systematics in the background subtraction dominate the uncertainties. The MDF curves show that a 5$\sigma$ detection of a 0.1~mCrab-like source takes about 200~s, while a signal-to-noise ratio of 100 (needed e.g., to investigate emission line profiles in X-ray spectra of accreting BHs and NSs) is reached in 5000~s for a 1~mCrab source. The background level and residual systematics requirements are driven by the strong field gravity objectives for relatively faint (1--10~mCrab) sources like most AGNs. Simulations show that, if the background variations are modelled on timescales of a few ks to a level of 0.3\% of the mean background level, it will be possible for example to observe a large sample of AGNs to measure BH spins with an accuracy of 20\%, and to carry out reverberation mapping measurements on AGNs \cite{DeRosa2018, uttleyetal14}.
\begin{table}[H]
\begin{center}
\caption{PFA main specifications.}
\label{tab:PFA}
\footnotesize
\begin{tabular}{l|l}
\bottomrule
\textbf{Parameter} & \textbf{Value} \\
\hline
Gas mixture & pure DME (CH3-O-CH3) at 0.8 atm\\
Absorption depth & 1 cm\\
Window & 50 $\mu$m Be\\
Energy range & 2–8 keV \\
Effective area & 915 cm$^{2}$~at 2~keV \\
& 495 cm$^{2}$~at 3~keV\\
& 216 cm$^{2}$~at 4~keV\\
& 46 cm$^{2}$~at 6~keV\\
Modulation factor & 38\% at 3 keV\\
& 57\% at 6 keV\\
Energy resolution (FWHM/$E$) & $<18\%$ at 6 keV \\
Field of View & 8 arcmin \\
Time resolution& $<500$~$\mu$s\\
\bottomrule
\end{tabular}
\end{center}
\end{table}
\subsection{Polarimetry Focusing Array -- PFA}
\label{sec:PFA}
The PFA consists of 4 identical telescopes optimized for X-ray imaging polarimetry, sensitive in the energy range of 2-8 keV. In synergy with the SFA and LAD, the PFA offers spatial, energy, and/or temporal resolved X-ray polarimetry at high sensitivity. PFA is also the only instrument on eXTP that has an imaging capability better than an arcminute, which may help to discriminate and remove source confusions occurring in other instruments. The main features of the PFA optics have already been discussed in section \ref{sec:SFA}. 
The PFA parameters and specifications are listed in Table \ref{tab:PFA}.

\begin{figure}[H]
\centering
\includegraphics[width=\columnwidth]{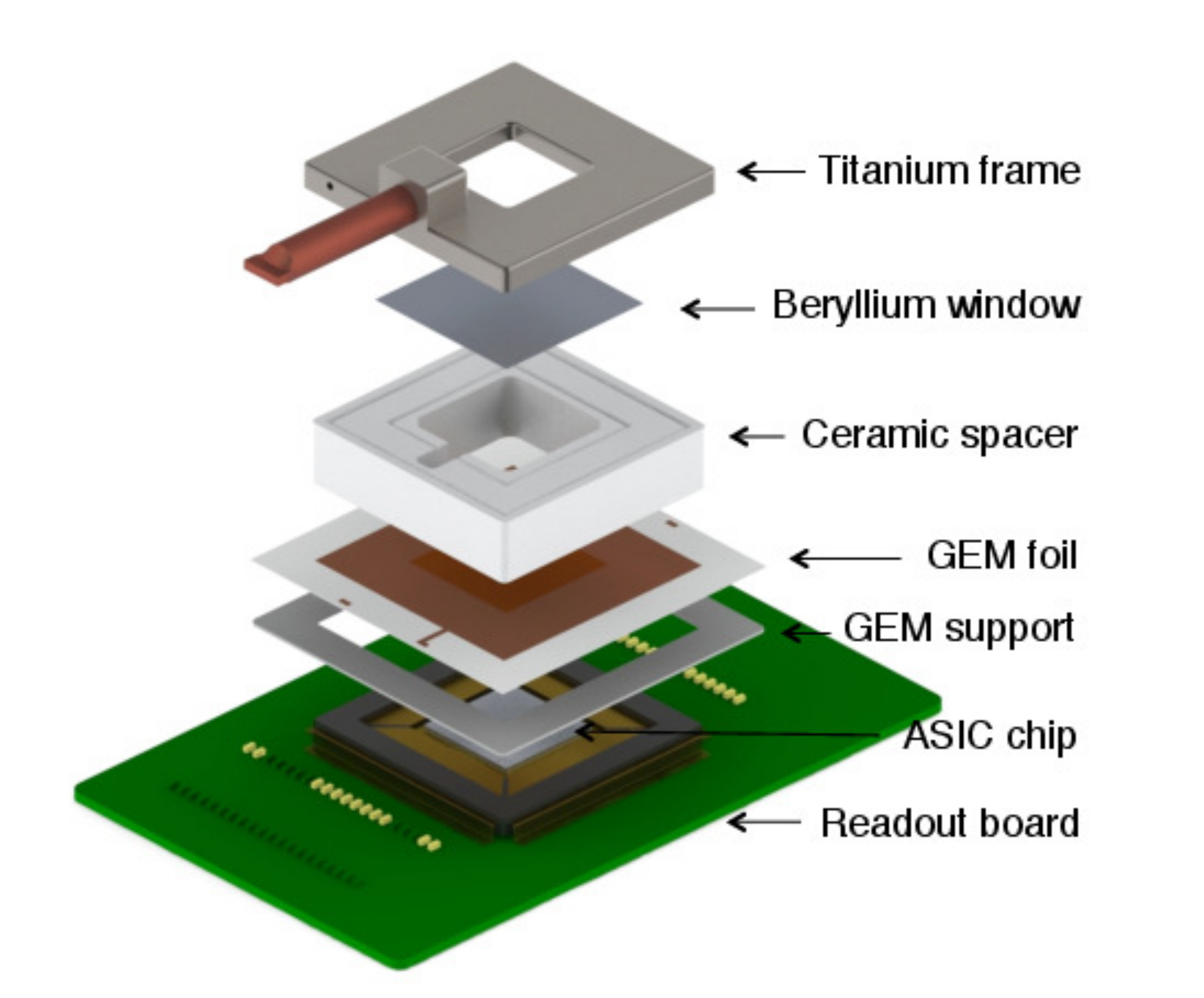}
\caption{A schematic drawing of the GPD.}
\label{fig:gpd}
\end{figure}

\subsubsection{Detectors and electronics.}
\label{sec:pfa:det}

\begin{figure}[H]
\includegraphics[width=0.49\columnwidth]{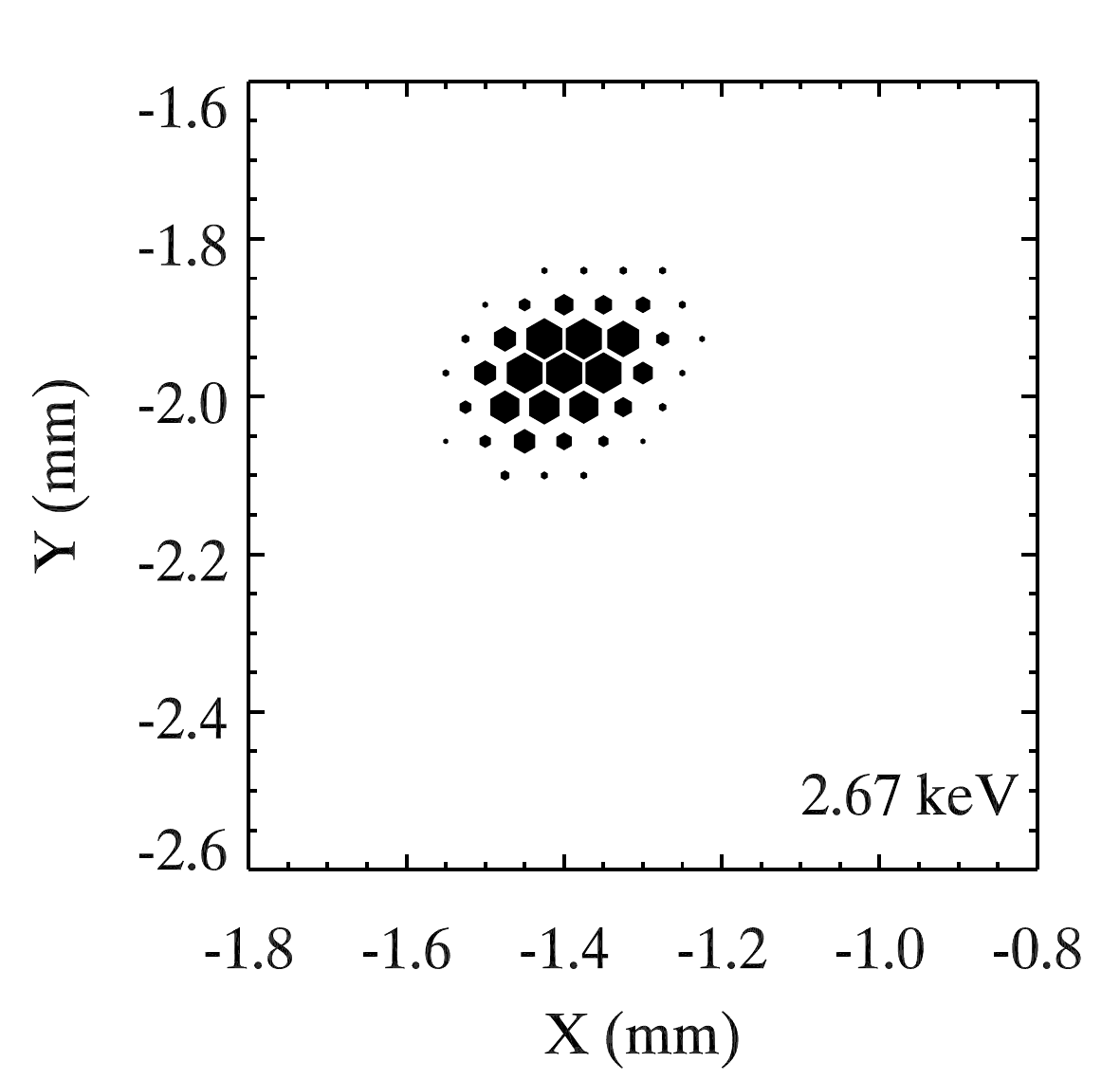}\\
\includegraphics[width=0.49\columnwidth]{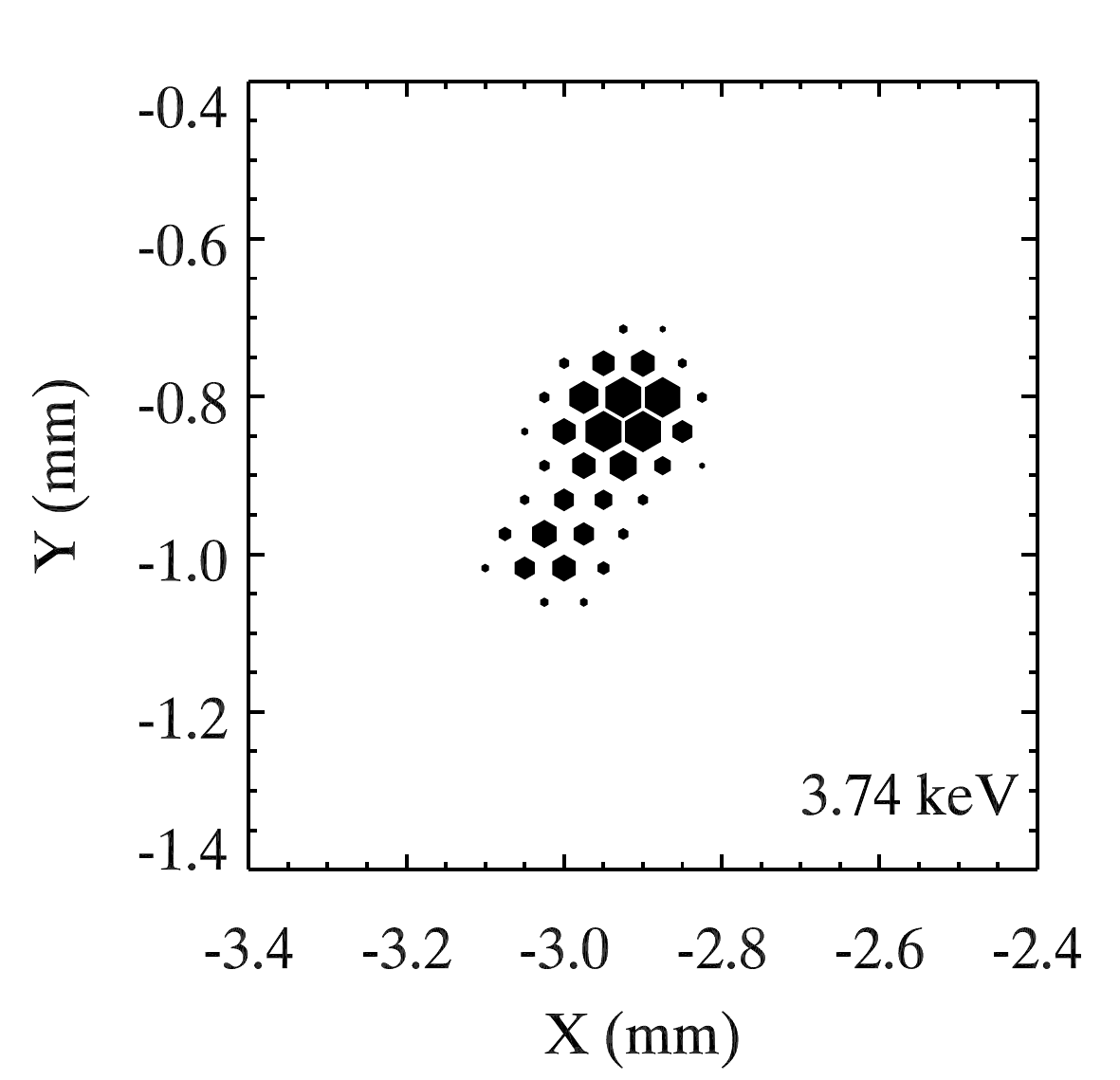}
\includegraphics[width=0.49\columnwidth]{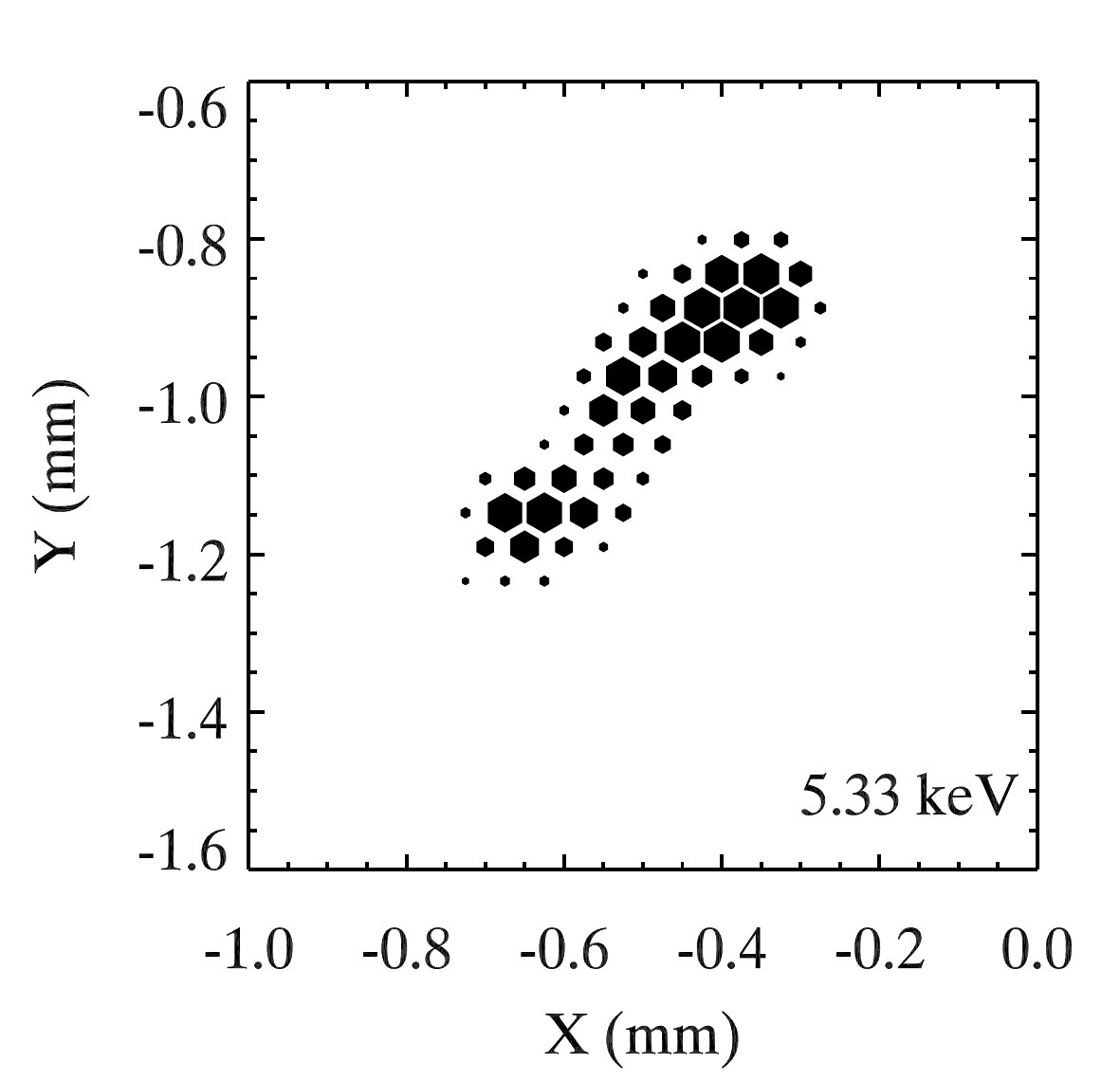}
\includegraphics[width=0.49\columnwidth]{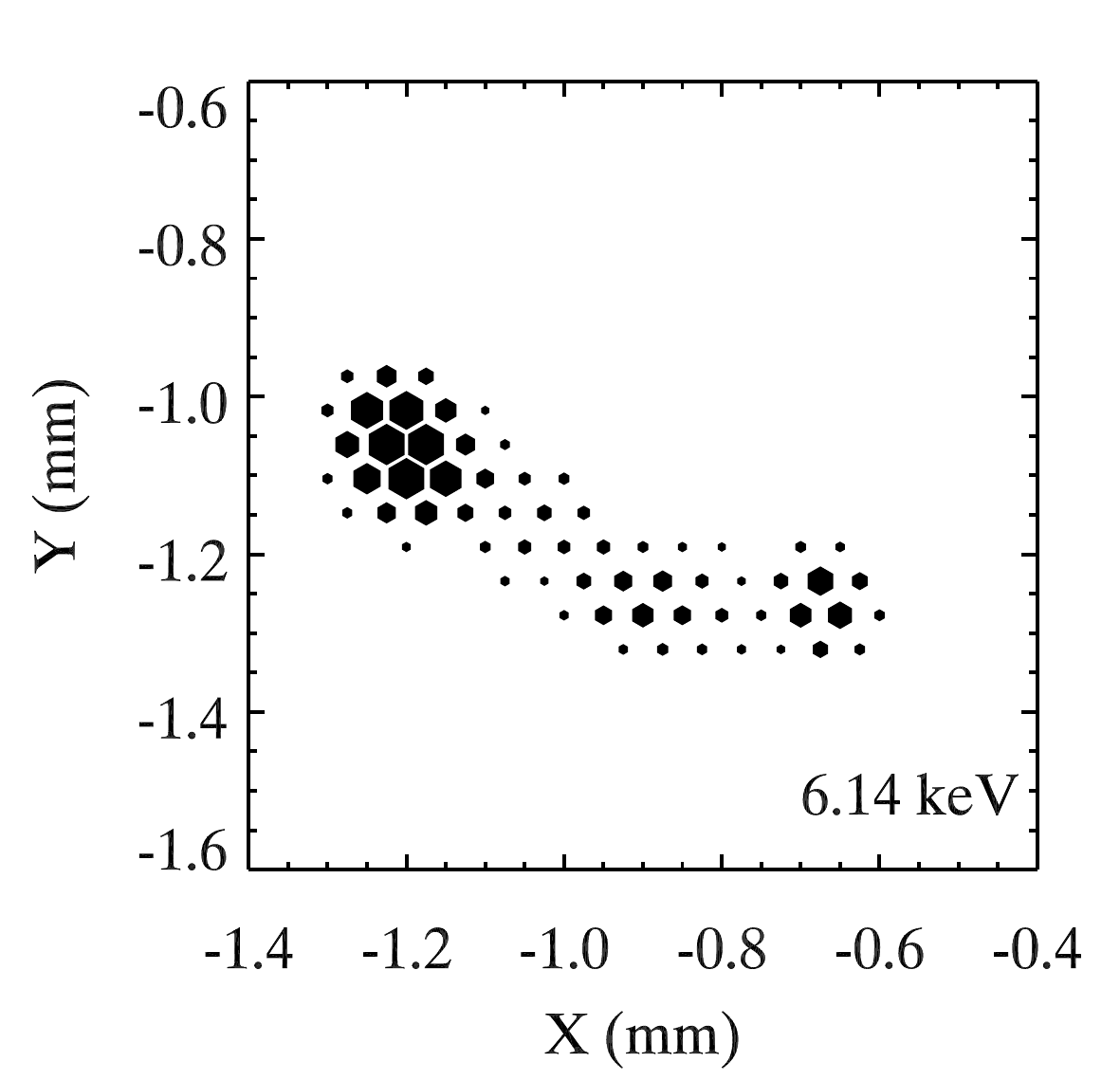}
\includegraphics[width=0.49\columnwidth]{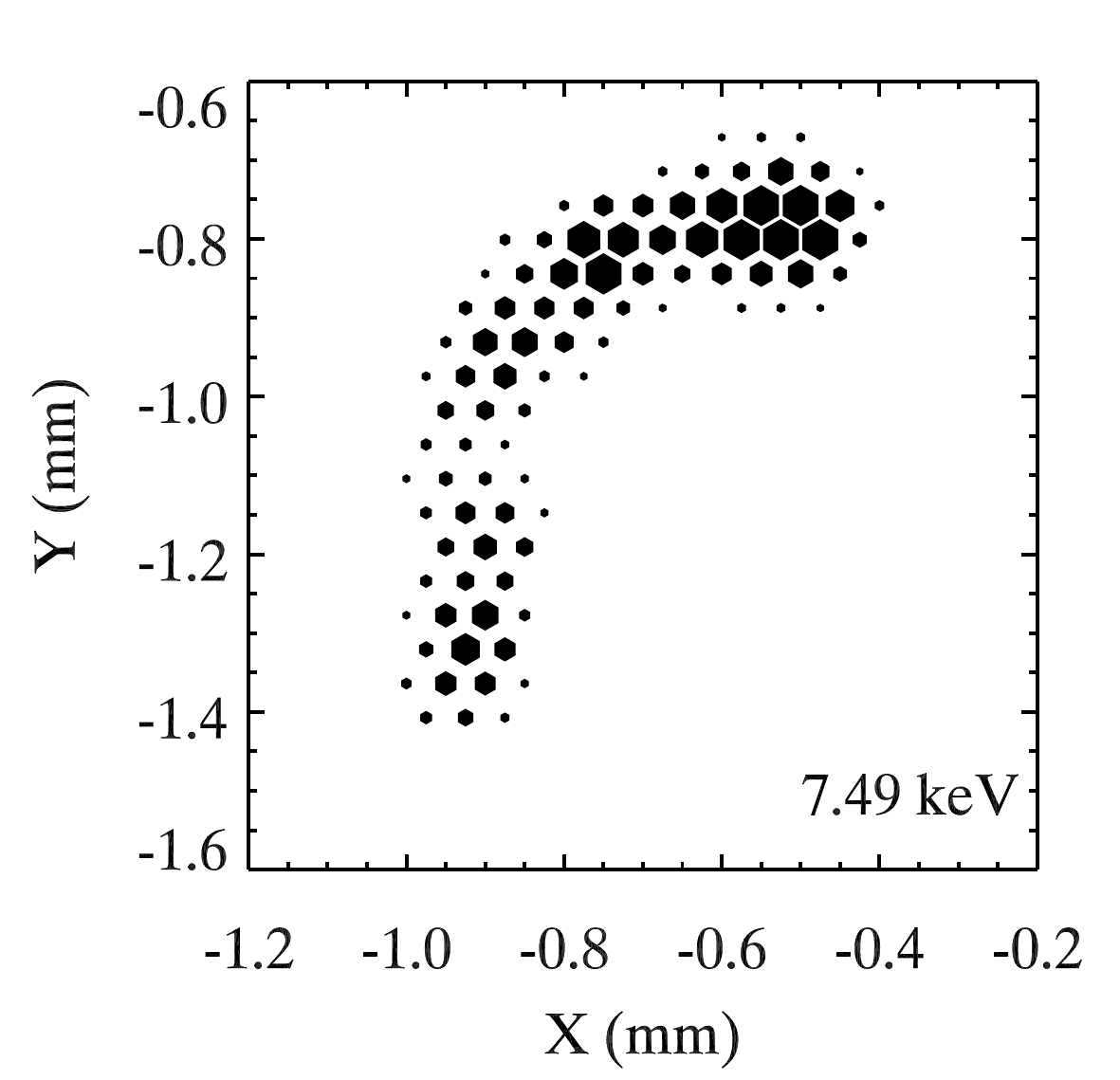}
\caption{Example track images measured with the GPD at different energies.}
\label{fig:track}
\end{figure}

The gas pixel detector (GPD) is adopted as the focal plane polarimeter for the PFA. The GPD was invented and developed by the INFN-Pisa group, and its principles are widely described in ~\cite{Costa2001,Bellazzini2003a,Bellazzini2007b,Bellazzini2013}. A schematic drawing of the detector is shown in Figure~\ref{fig:gpd} for illustration. It is a gas chamber sealed by a 50~$\mu$m thick Beryllium entrance window. The chamber is vacuumed and baked out before filled with the working gas (dimethyl ether or DME at 0.8 atm), which absorbs the incident X-rays and converts them to photoelectrons, whose emission angle encodes the information about the X-ray polarization and is thus the key physical parameter to be measured. A drift field of about 2~kV~cm$^{-1}$ is applied in the chamber to drive secondary electrons ionized by the initial photoelectron to move toward the anode. 
To enable measurements with a sufficient signal-to-noise ratio, a gas electron multiplier (GEM) is mounted above the anode to multiply the number of electrons by a gain factor of a few hundred, which can be adjusted by the high voltage across the top and bottom layer of the GEM.  Beneath the GEM is positioned the key element of the GPD, the readout ASIC chip~\cite{Bellazzini2004, Bellazzini2006b}, which is pixelated to have a pitch of 50~$\mu$m and responsible for the collection and measure of the charges after multiplication. The ASIC has a dimension 1.5 cm $\times$ 1.5 cm, which defines the sensitive region of the detector. The readout noise is around 50~e$^-$ for the ASIC. The energy resolution is typical for a gas detector, i.e., 15-20\% at 6~keV.  In brief, with the GPD, one is able to measure the 2D ionization track of the photoelectron in the gas chamber (see Figure~\ref{fig:track} for examples), and infer the polarization of the incident X-ray beam via the modulation of the emission angle reconstructed from the track image. The BEE is designed to control and operate the ASIC, drive the analog to digital conversion, organize and store the data, and communicate with the satellite. They also regulate the high voltage modules, which are needed for the drift field and to power the GEM field.

\subsubsection{Performance: Area, Background, MDP.}
\label{sec:pfa:perf}

\begin{figure}[H]
\centering
\includegraphics[width=\columnwidth]{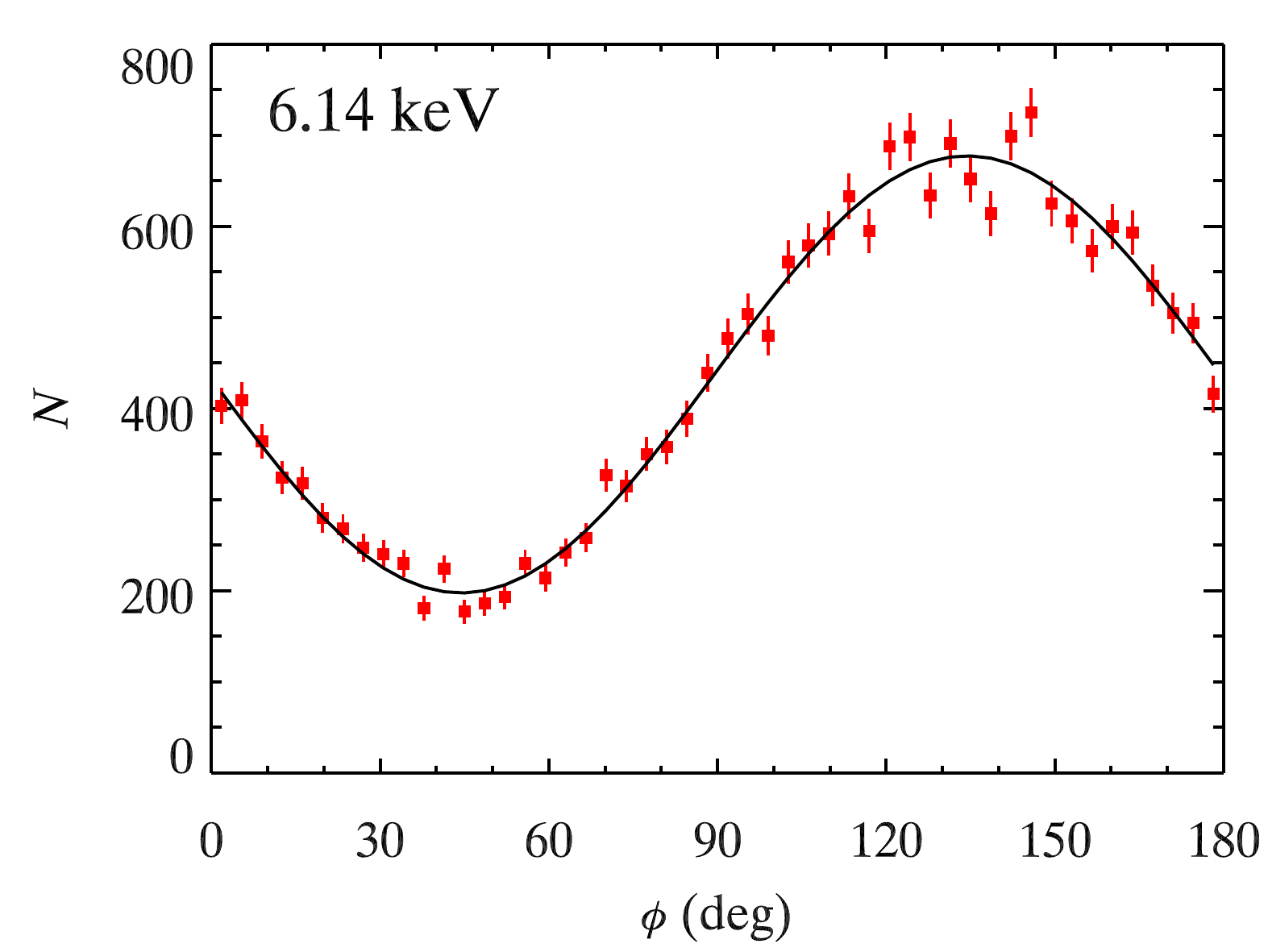}
\caption{Modulation curve measured with fully polarized X-rays of 6.14 keV (45$^\circ$\ Bragg diffraction with a LiF crystal). The curve gives the best-fit modulation function, with a degree of modulation $58.2\% \pm 0.7\%$.}
\label{fig:modcurve}
\end{figure}

\begin{figure}[H]
\centering
\includegraphics[width=\columnwidth]{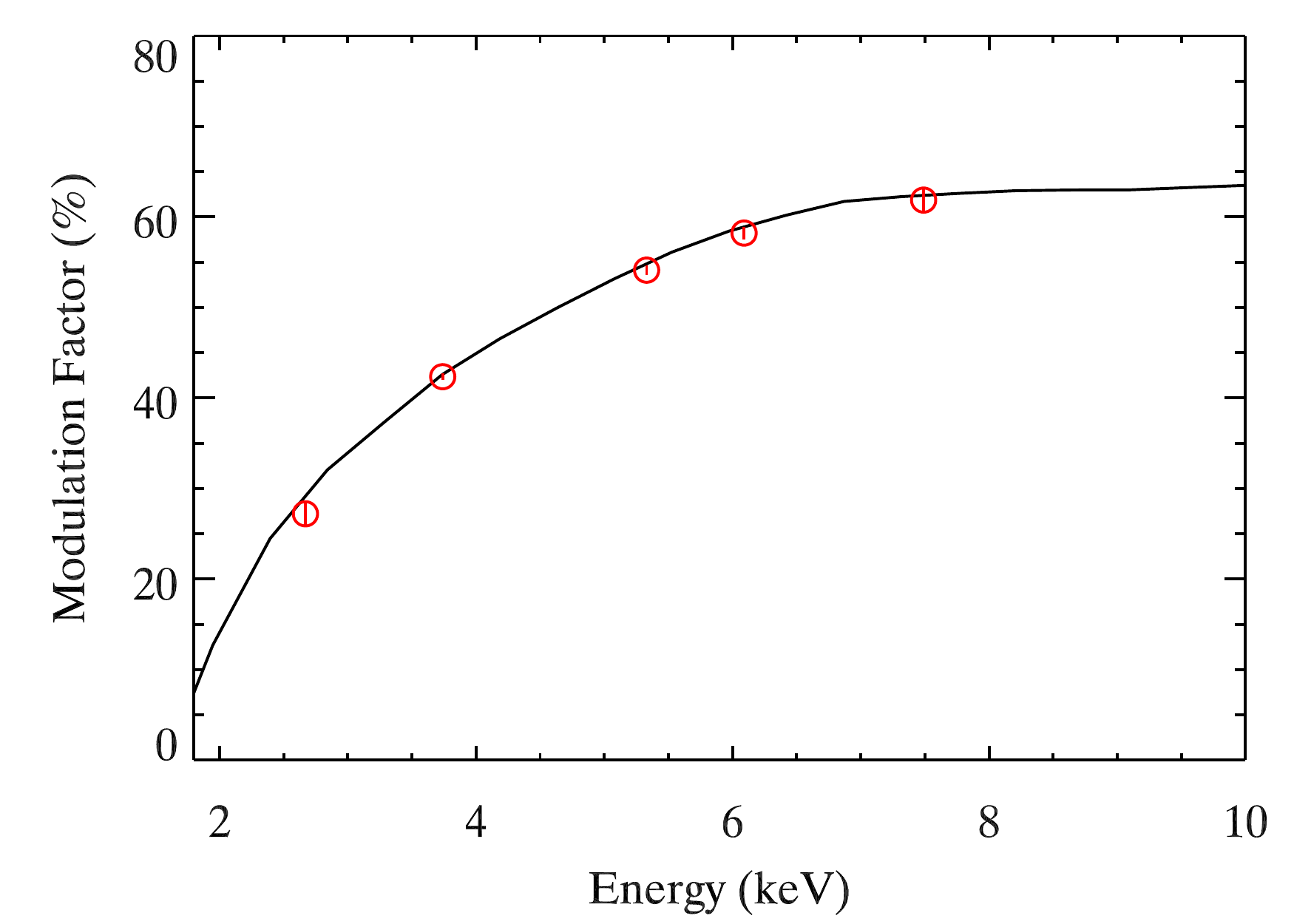}
\caption{Modulation factor of the GPD. The points are measurements and the curve is obtained from simulation.}
\label{fig:mod}
\end{figure}

For a polarized X-ray source, the distribution of the photoelectron emission angle ($\phi$) projected on the focal plane is modulated by the cosine function (see Figure~\ref{fig:modcurve} for an example) $N(\phi) = A + B \cos^2(\phi - \phi_0)$,
where $A$ and $B$ are constants and $\phi_0$ is the position angle of polarization. The amplitude of the modulation is linearly scaled with the degree of polarization of the incident beam. In response to a fully polarized X-ray source, the amplitude of modulation is also called the modulation factor $\mu = \frac{\textmd{max} - \textmd{min}}{\textmd{max} + \textmd{min}} = \frac{B}{2A + B}$. This is one of the most important parameters that determine the sensitivity of the polarimeter. The simulated and measured modulation factor versus energy is shown in Figure~\ref{fig:mod}. It decreases with decreasing energy for two reasons. First, the Coulomb scattering of the photoelectron by the nucleus will randomize the emission angle and lower the modulation degree, which becomes more important at low energies. Second, due to the shorter range of the track at low energies, the reconstruction of the emission angle becomes more difficult and is limited by the finite pixel size.

\begin{figure}[H]
\centering
\includegraphics[width=\columnwidth]{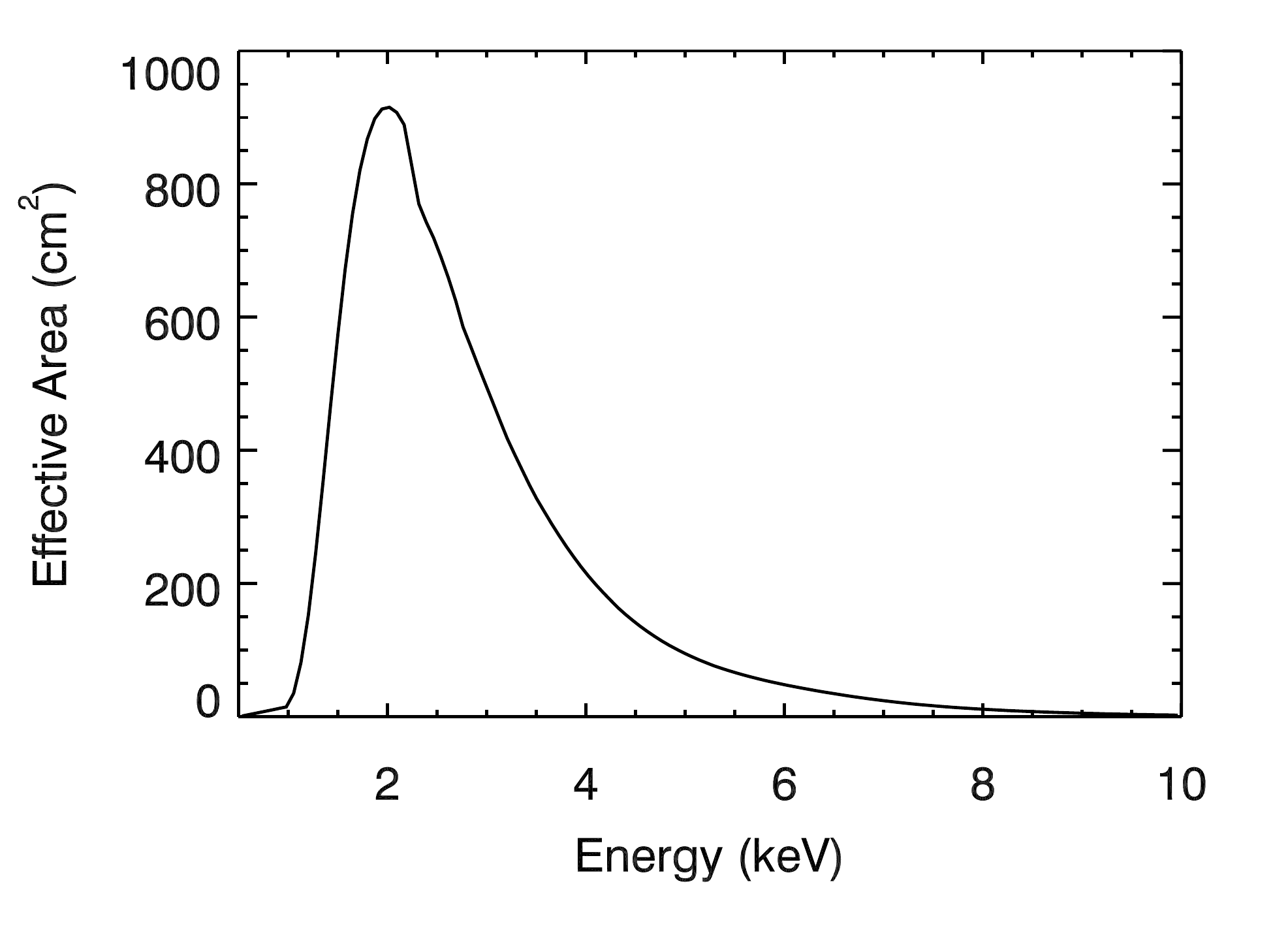}
\caption{Total effective area of the four PFA telescopes in combination with the optics and detector.}
\label{fig:effarea}
\end{figure}

The total effective area of the four telescopes, and including the GPD, is shown in Figure~\ref{fig:effarea} summed for the four telescopes in PFA. The total background, taking into account the cosmic diffuse X-ray background and the particle induced background, is estimated to be about $6 \times 10^{-3}$~counts~s$^{-1}$ in the energy range of 2-8~keV in the source aperture. By comparison, the Crab nebula will result in $\sim$1,200~counts~s$^{-1}$ with the 4 GPDs in the same energy band. As the majority of targets of eXTP will be bright sources in our Milky Way, such a background will be negligible for observations in most cases. 

\begin{figure}[H]
\centering
\includegraphics[width=\columnwidth]{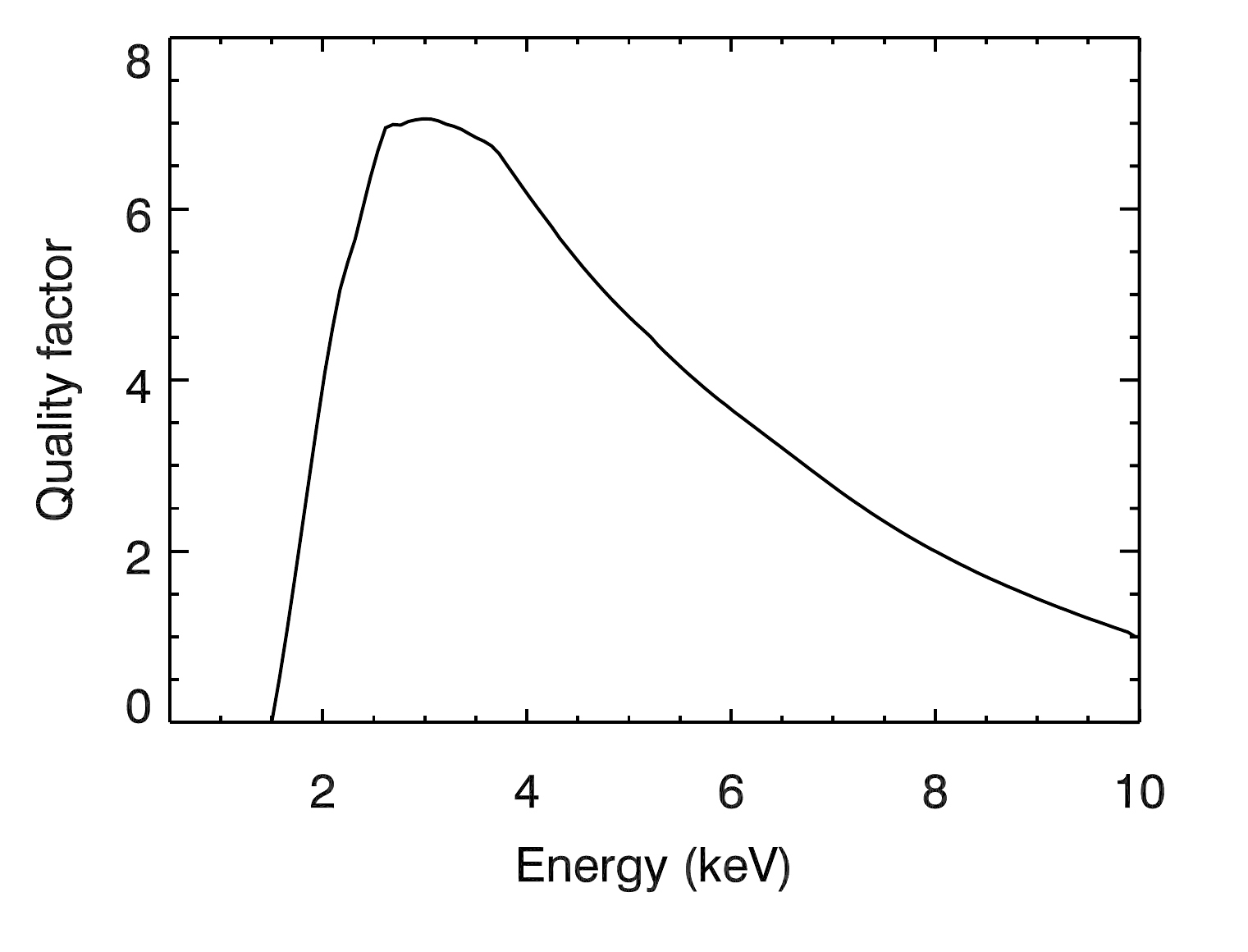}
\caption{Quality curve ($\mu \sqrt{A}$) for the four PFA telescopes.}
\label{fig:quality}
\end{figure}

The sensitivity of the polarimeter is usually defined as the minimum detectable polarization (MDP), described as
$\textmd{MDP} = \frac{4.29}{\mu 
} \sqrt{\frac{SA + B}{t}}$, where $\mu$ is the modulation factor, $A$ is the effective area, $S$ is the source intensity in photons~cm$^{-2}$~s$^{-1}$, $B$ is the background rate in the source aperture in counts~s$^{-1}$, $t$ is the exposure time, and 4.29 corresponds to a confidence level of 99\%.  In cases where the background is not important,  the MDP can be approximated as $\textmd{MDP} = \frac{4.29}{\mu \sqrt{SAt}}$.
Thus, the sensitivity is not only a function of the effective area, but correlated with $\mu \sqrt{A}$, the quality curve of a polarimeter. Given the modulation factor and the effective area, the quality curve of PFA is shown in Figure~\ref{fig:quality}. As one can see, the effective area peaks at about 2~keV while the sensitivity peaks at around 3~keV due to the increase of the modulation factor with energy. 

Another important parameter for a polarimeter is its systematic error, i.e., the residual modulation seen with fully unpolarized sources. The systematic error will limit the sensitivity of the polarimeter no matter how bright the source is and how long the exposure is.  The low systematics of the GPD have been widely discussed in~\cite{Muleri2010,Muleri2012,Bellazzini2013,Li2015}. In general, the systematics can be controlled below 1\% with the GPD only. The optics is not expected to contribute to the systematics of polarization measurements due to its grazing incidence angle and symmetric geometry. However, this needs to be tested in the future.  If the instrument is well calibrated, the systematics can be determined and subtracted, leading to rather low systematics. 

A simple way to estimate the sensitivity given a source with a spectrum similar to that of the Crab nebula is described below. In the energy band of 2-8~keV, the mean modulation factor is $\sim$0.23 weighted by the observed Crab nebula spectrum. In the case of negligible background, an exposure of 1~ks of the Crab nebula will result in an MDP of 1.7\%. Sensitivities for observations with sources of different intensities and different exposure times can be scaled by simply using the equation for the MDP. We also note that the MDP quoted above indicates a detection at 99\% confidence level (a chance probability of 1\% of having such a level of measurement from a  fully unpolarized source), and one needs more observing time to achieve a precise measurement with a significance of 3$\sigma$ or more~\cite{Strohmayer2013}.

\begin{figure}[H]
\centering
\includegraphics[width=\columnwidth]{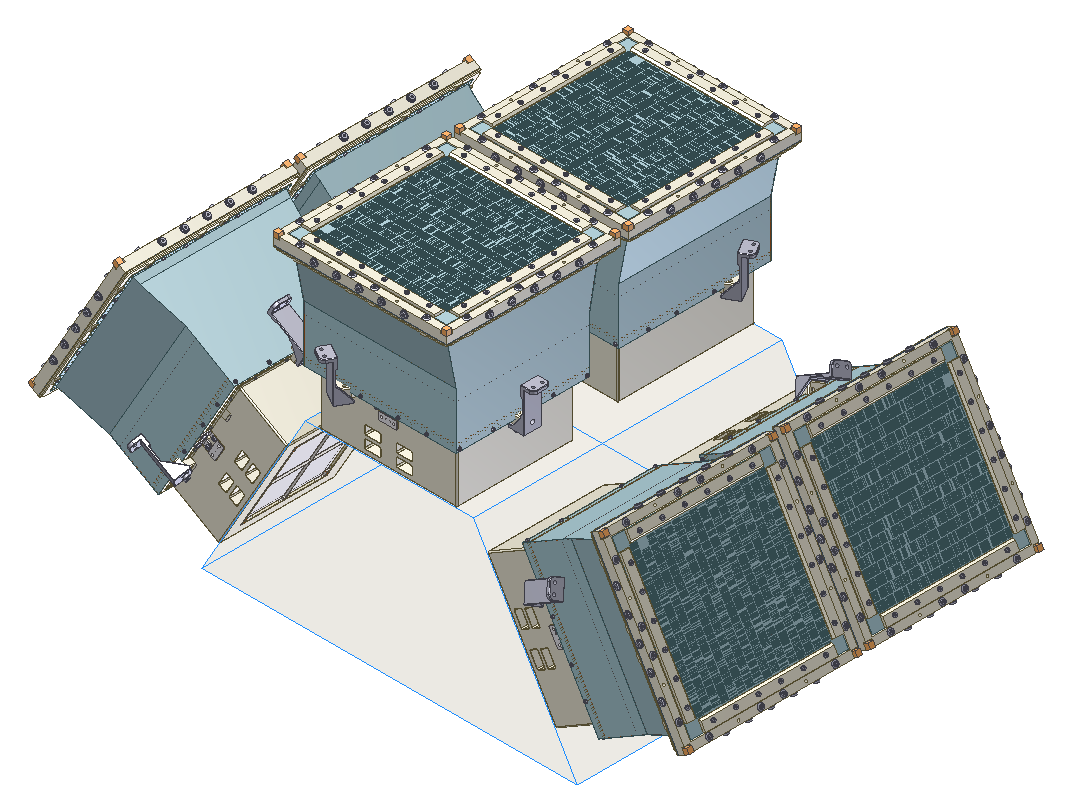}
\caption{WFM configuration based on three camera pairs. Each camera pair is built by two orthogonal WFM cameras.}
\label{fig:WFM}
\end{figure}

\subsection{Wide Field Monitor -- WFM}
 \label{sec:WFM}

The WFM is based on three pairs of coded mask cameras equipped with position-sensitive SDDs, covering $4.1$~sr ($\sim 33$\%) of the sky and operating in the energy range $2-50$~keV \cite{Brandt2014}. The effective FoV of each camera pair is $\sim 70^{\circ} \times 70^{\circ}$ and $\sim 90^{\circ} \times 90^{\circ}$ at zero response. The peak sensitivity is reached along the LAD pointing direction. The energy resolution is $\sim300$~eV at  6~keV, and the absolute time accuracy is 1~$\mu$s. 
Pairs of two orthogonal cameras are combined to obtain accurate 2D positions of the monitored sources. The three camera pairs (six cameras in total) are arranged in a baseline configuration of $-60^{\circ}$, $0^{\circ}$, $60^{\circ}$ as shown in Figure~\ref{fig:WFM}. The WFM main parameters and anticipated performance are listed in Table \ref{tabWFM}. 

\begin{figure}[H]
\centering
\includegraphics[width=0.9\columnwidth]{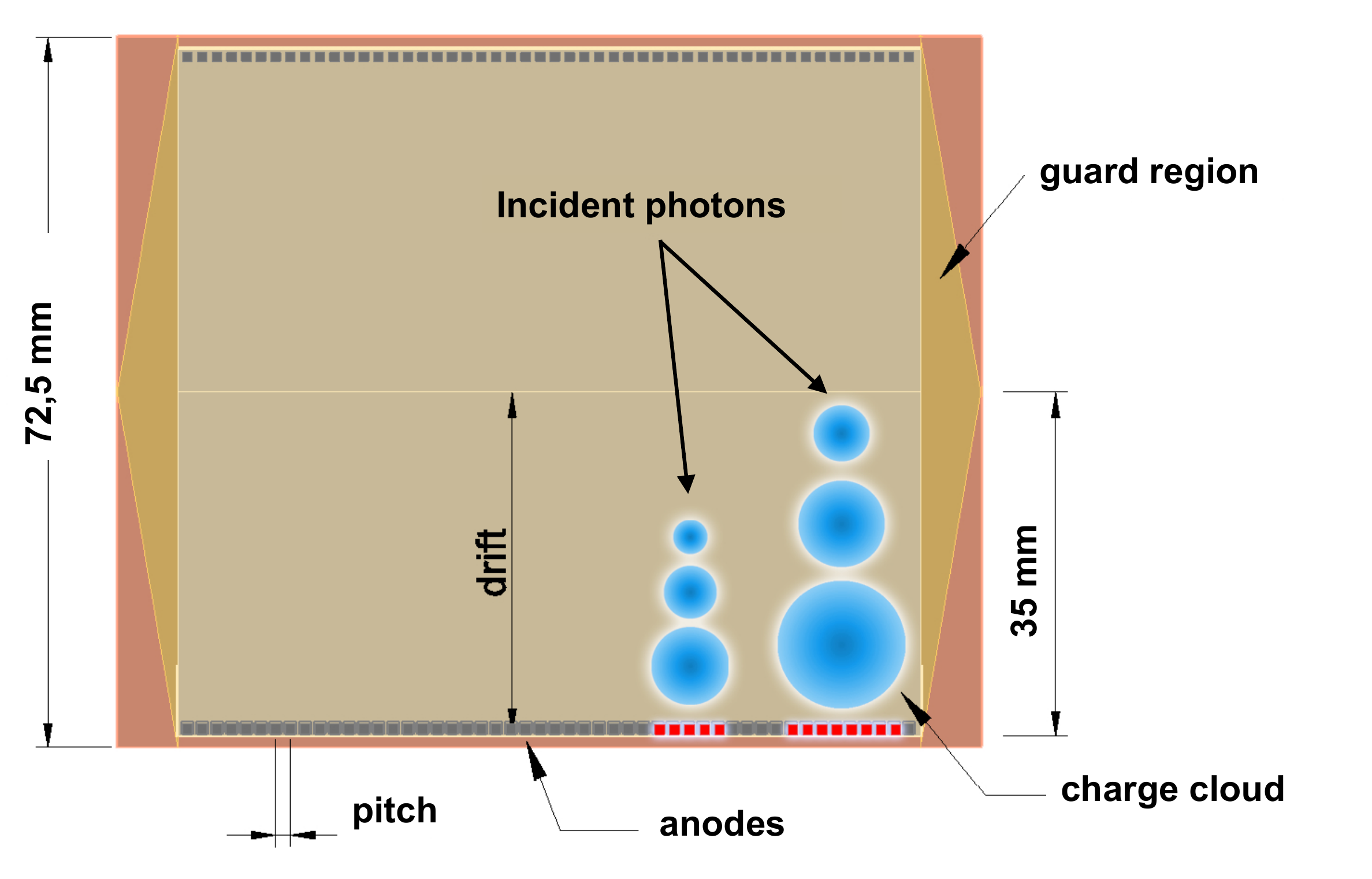}
\caption{Details of the Working Principle.}
\label{fig:WFM_working}
\end{figure}

\begin{figure}[H]
\centering
\includegraphics[width=0.9\columnwidth]{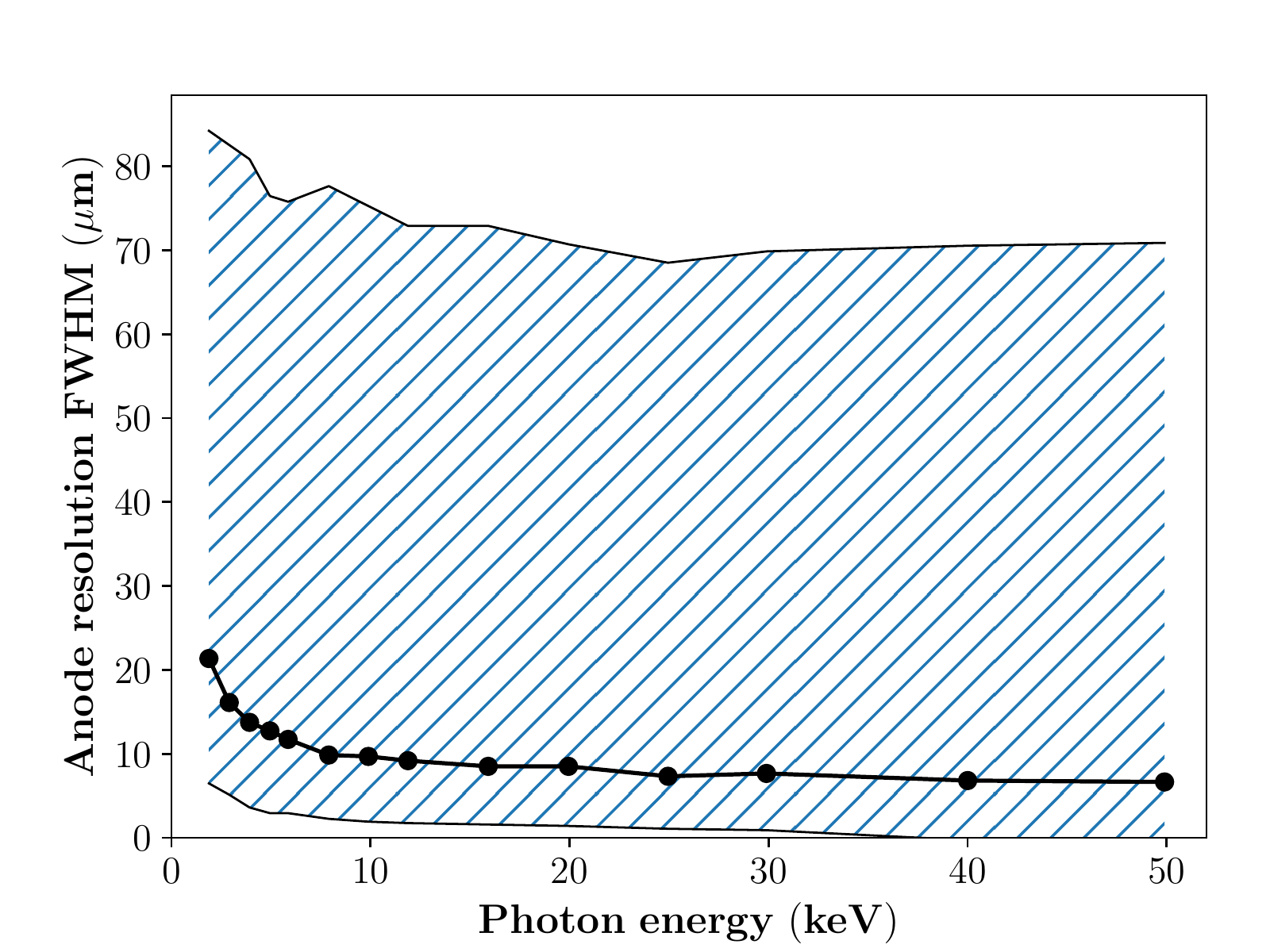}
\includegraphics[width=0.9\columnwidth]{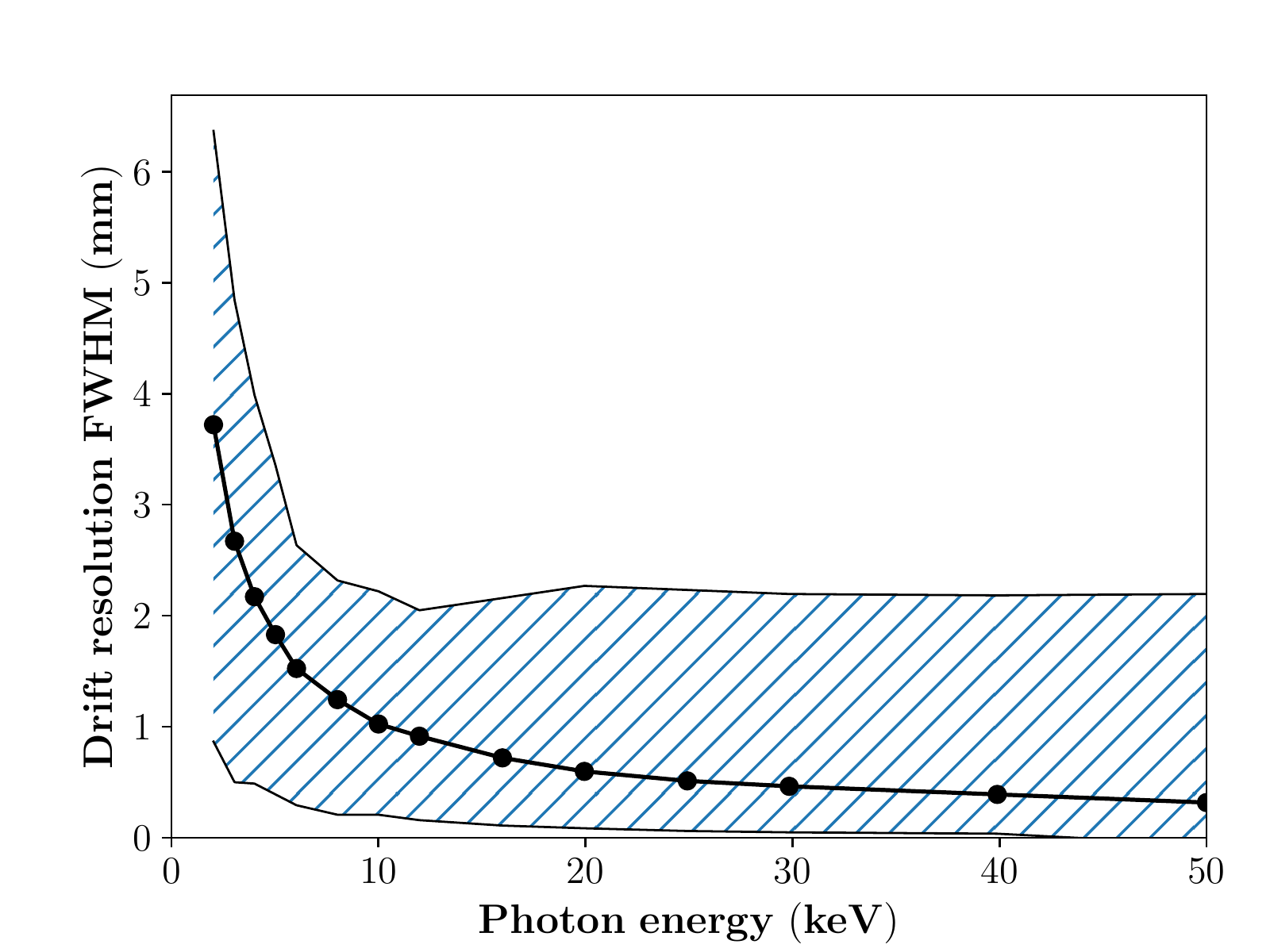}
\caption{SDD anodic (top) and drift (bottom) spatial resolution (FWHM) as a function of the photon energy for the WFM detectors. The shaded area represents minimum-maximum spatial resolution performance inside the detector channel. Filled circles show the averaged channel values.}
\label{fig:WFM_work1}
\end{figure}

The WFM detector plane is based on the same large area SDD technology developed for the LAD. The detector geometry has been sligthly modified to enable 2D imaging. When a photon is absorbed by the SDD, it generates an electron cloud that is focused on the middle plane of the detector, and then drifts towards the anodes at constant speed. As already discussed in section \ref{sec:lad_detectors_and_electronics}, while drifting, the electron cloud size increases due to diffusion. As we already said, it can be described with a Gaussian function, with an area $A$ equal to the total charge (i.e the photon energy), a mean value $m$ representing the "anodic" coordinate of the impact point and a width $\sigma$, which depends on the "drift" coordinate of the absorption point. See also Figure \ref{fig:WFM_working}.

\begin{table}[H]
\begin{center}
\caption{The WFM main parameters and performance.}
\label{tabWFM}
\footnotesize
\begin{tabular}{l|l}
\bottomrule
\textbf{Parameters} & \textbf{Anticipated value} \\
\hline
Location Accuracy & $<1$~arcmin\\
&$<30$~arcsec for P/L\\
Angular resolution  & $<4.3$~arcmin (FWHM)\\
& req. $5$~arcmin (FWHM)\\
Peak sensitivity in LAD direction & $<0.6$~Crab (1s), req. $<1$~Crab\\
& 2.1~mCrab (50 ks), req. 5~mCrab\\
Absolute flux accuracy & $<20$\%\\
Field of view at 0\% response & $1.75{\pi}$~sr=5.5~sr \\
at $20$\% of peak camera response & $1.33{\pi}$~sr =4.1~sr \\
Energy range & 2--50~keV \\
Energy resolution (FWHM) & $<300$~eV at 6~keV \\
 & req. $<500$~eV at 6~keV \\
Energy scale accuracy & $< 2$\%\\
Energy bands for compressed images & $\geq  64$ \\
Field of view & Camera: $\sim 90^{\circ} \times 90^{\circ}$ FWZR\\
& 3 Pairs $\sim 180^{\circ} \times 90^{\circ}$ FWZR\\
Time resolution & $\leq 300$~s for images\\
& $<10{\mu}$s for event data\\
Absolute time calibration & $<2{\mu}$s, $1~{\mu}$s for P/L\\
Burst Trigger scale &10~ms up to 300~s\\ 
Number of GRB Triggers & $>1$ GRB triggers per orbit\\
Mass & 11.2~kg per camera\\
& 78.5~kg for 3 pairs (including ICU)  \\
Nominal Power & 72 W (including ICU) \\
Detector Op. Temperature & $<-20^{\circ}$~C \\
Typical/Max data rate & 50/100 kbits/s (after compression)\\
\bottomrule
\end{tabular}
\end{center}
\end{table}

\begin{figure}[H]
\centering
\includegraphics[width=\columnwidth]{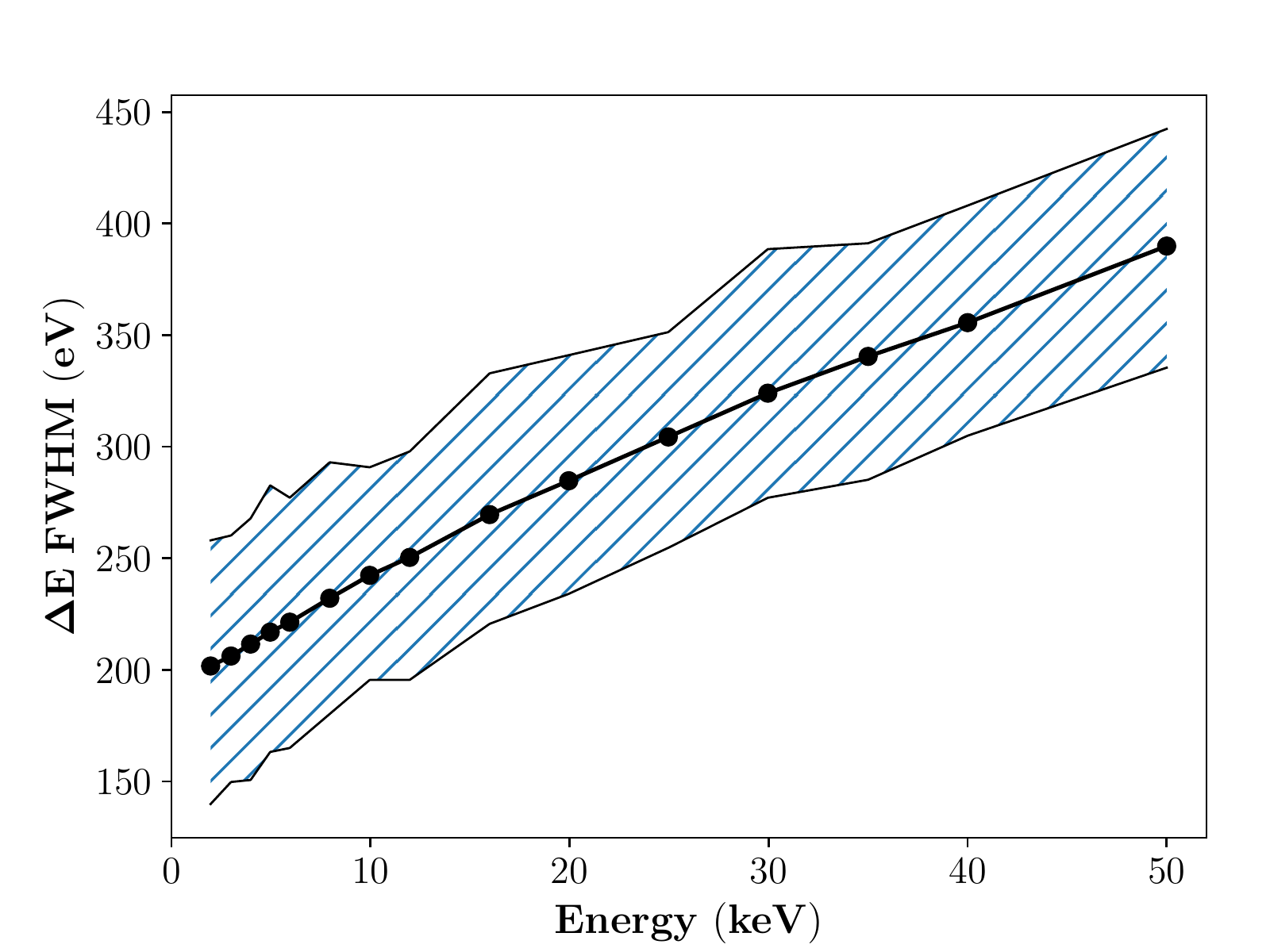}
\caption{Energy resolution (FWHM) of the WFM detectors as a function of the photon energy. Solid circles represent the resolution values averaged on the whole SDD channel while the shaded area shows the range of energy resolution for each photon energy.}
\label{fig:WFM_work2}
\end{figure}

For each photon, we measure the energy, proportional to the collected charge, the so-called X-position, the center of the charge cloud ($<60~\mu$m), the Y-position proportional to the width of the charge cloud ($<8$~mm), and the time of the event. For each event the analysis of the charge distribution over the anodes is performed on-board by the FPGA-based BEE, and provides the amplitude, the anode and drift positions of the event. In order to maximize the signal-to-noise ratio of the anode and drift position information, the design of the WFM SDD has been optimized by means of Monte Carlo simulations \cite{Evangelista2014}. This resulted in a smaller anode pitch with respect to the LAD one (169~$\mu$m vs 970~$\mu$m) and quasi-squared overall dimensions ($7.74 \times7.25$~cm$^2$ geometric area, $6.50 \times7.00$~cm$^2$ effective area).
Taking into account the eXTP radiation and thermal environments, the relevant parameters for the detector and read-out electronics, and the current choice of the anode pitch, the expected position and energy resolution have been estimated and are shown in Figure~\ref{fig:WFM_work1} and Figure~\ref{fig:WFM_work2}.

 \subsubsection{The WFM architecture}
\begin{figure}[H]
\centering
\includegraphics[width=\columnwidth]{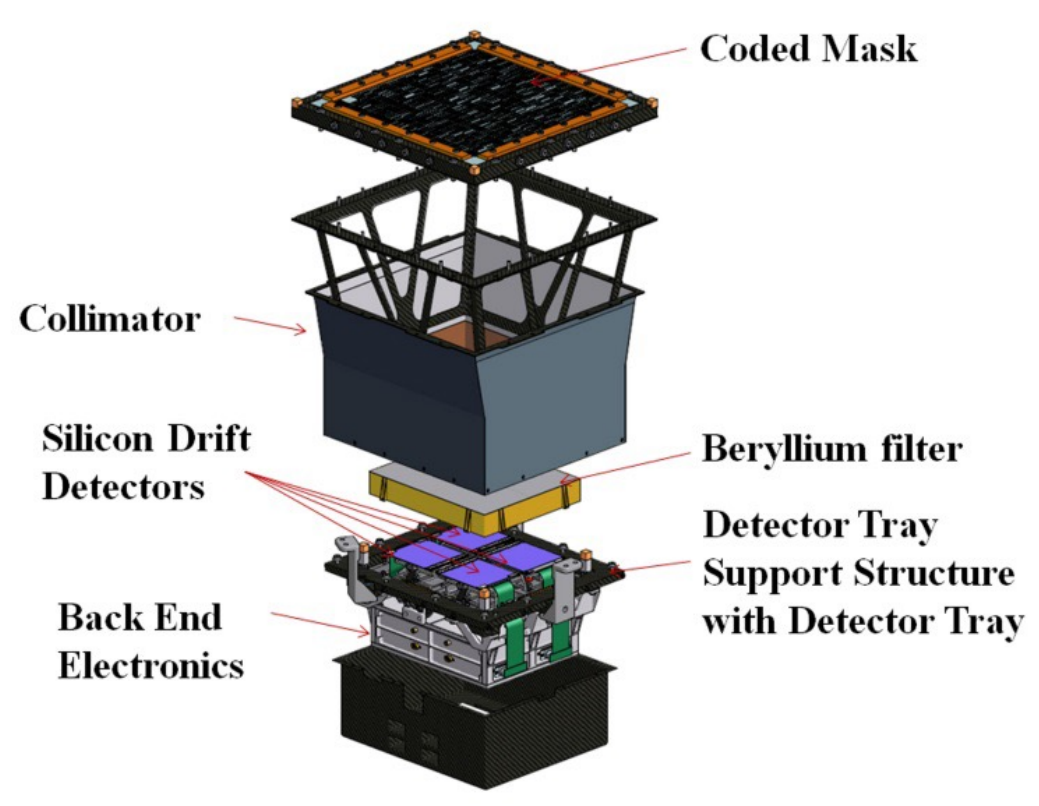}
\caption{Main elements of the WFM camera design}
\label{fig:WFM_exploded}
\end{figure}

The main elements of the WFM are shown in Figure~\ref{fig:WFM_exploded}. 
The WFM instrument includes three camera pairs, that is six identical and independent cameras, each one with its own BEE, and two ICUs for cold redundancy. Each camera is composed of one detector tray with four SDDs, four FEEs, four Be windows, one BEE assembly, one  collimator, and one coded mask with a thermal blanket as shown in Figure~\ref{fig:WFM_exploded}. The camera pairs are designed with a high level of redundancy. The mass and power resources are listed in Table \ref{tabWFM}.

 \textit{The WFM optical design}. The WFM is a coded mask instrument: photons from a certain direction in the sky project the mask pattern on the position sensitive SDD. The coded mask has a pattern consisting of 1040 x 16 open/closed elements. The mask pitch is 250~$\mu$m (fine resolution) $\times$~16.4~mm (coarse resolution). The dimensions of the open elements are 250~$\mu$m $\times$ 14~mm with 2.4~mm spacing between the elements in the coarse resolution direction for mechanical reasons. The mask nominal open fraction is chosen to be $\sim25\%$ to optimize the sensitivity for weaker sources. The detector-mask distance is 202.9~mm. This assures the required angular resolution (expressed in FWHM), which for the on-axis viewing direction is Arctan (p/d), where p is the mask pitch and d is the distance between the detector plane and the mask. Thus the angular resolution in the direction of the fine mask pitch is 4.24~arcmin and in the coarse direction it is 4.5~deg. The size of the mask is $\sim$1.7 times larger than the detector plane to achieve a flat (i.e., fully illuminated) region in the center of the FoV. The fine position resolution in the two coordinates is obtained by combining the images produced by the two orthogonal and co-aligned cameras forming each WFM camera pair.

\textit{Coded mask, mask frame}. The coded mask area is 260~mm $\times$ 260~mm and is manufactured from a 150~$\mu$m thick Tungsten foil. There are strict requirements on the flatness and stability of the coded mask. The mask must be flat (or at least must maintain its shape) to $\pm50$~$\mu$m over its entire surface across the full operational temperature range. Temperature gradients during the orbit must be less than $10^{\circ}$C. This is the main reason why a sunshield is a requirement for the WFM. The mask frame acts as a pretension mechanism in order to minimize the vertical displacements of the mask during the operational mode.

\textit{Collimator.} The collimator supports the coded mask frame assembly and is made of CFRP material. It has been selected because it is light and has enough stiffness to avoid deformations, which can appear during launch (accelerations) and operation (thermal stresses). The collimator will be covered by a Tungsten sheet as a background shield. In addition, Copper and Molybdenum plates will be placed in the inner part of the collimator for in-flight calibration purposes.

\textit{Detector tray.}
The detector tray assembly consists of four detector assemblies (DAs) and the detector support plate (DSP). Each detector assembly consists of an SDD tile glued on a ceramic PCB, containing the FEE. Three invar mounting legs provide the mechanical interface for alignment of the DA to the DSP. Likewise, DSP will be used to mechanically align the four DAs with respect to each other.

\textit{Beryllium window and detector tray support structure.}
A 25$\mu$m thick Beryllium foil is located above each SDD as protection against micrometeoroid impacts and debris.
The detector tray support structure (DTSS) serves as a support for the DSP, and facilitates the mounting of the collimator and BEE box support structure, which accommodates the BEE box. In addition, the isostatic fitting is mounted on the DTSS and serves as mechanical interface between the camera and the support structure.

\textit{Back-End Electronics and ICU.}
There is only one level of BEE, as there are only four detectors to be read out per camera. The BEE box is located along with the PSU at the bottom of each camera. Although most of the processing is similar to that of the LAD BEE, the WFM BEE has the additional capability to determine rather accurate photon positions for each photon from the FEE data. A higher level of processing power is therefore required inside the BEEs. High-resolution detector images in several energy bands are accumulated directly on board. For this purpose, the WFM BEE will be based on an RTG4 FPGA that is larger, faster and more flexible than the RTAX-SL. As in the PBEE of the LAD, the WFM BEE will transfer the science data products to the WFM ICU along with the housekeeping data via a SpaceWire interface.
The WFM ICU consists of a main and a redundant unit, housed in two separate boxes. The ICU controls each of the six cameras independently and interfaces with the PDHU, performing on-board computations to locate bright transient events in real time. 

\textit{Thermal control.} A stable thermal environment is obtained by protecting the instrument with a sunshield. The operating temperature range of the four SDD/FEE sandwiches are between $-30$ $^{\circ}$C and $-3$ $^{\circ}$C. Their internal heat dissipation will be dissipated mainly via conduction to the thermal interface provided by the S/C. The coded mask is covered by a thermal blanket in order to reduce mask temperature variations along the orbit and within the mask. The power dissipated by the BEE will be radiatively transferred to the optical bench and deep sky via the backside of the BEE box. In addition, an MLI blanket will wrap the collimator, to provide an optimal isolation to the SDD/FEE sandwich, and to keep the camera alignment within the desired margin during the mission.

\subsubsection{Performance.} \label{sec:WFM_Performance}
\begin{figure}[H]
\centering
\includegraphics[width=\columnwidth]{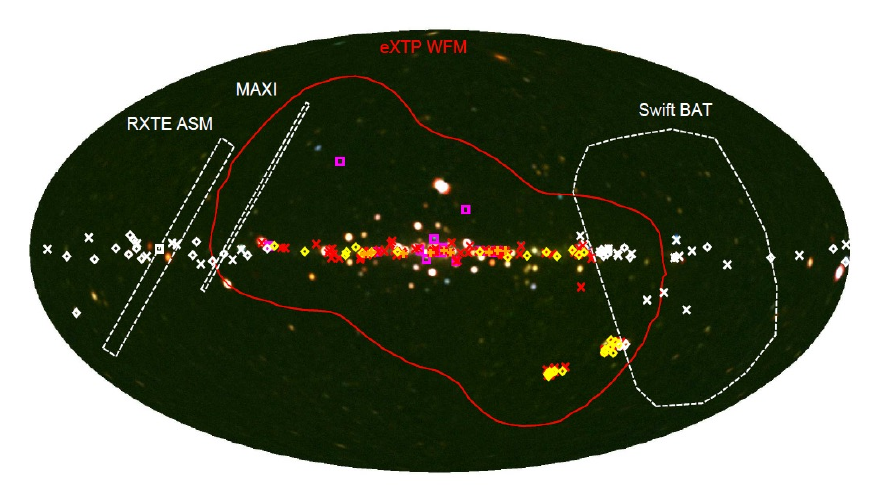}
\caption{The WFM FoV, red curve, is compared with other relevant all sky monitoring instruments. On the map, yellow diamonds mark the known Be X-ray binary systems, red crosses mark supergiant HMXBs, orange crosses the SFXTs, and magenta boxes the SyXBs. Grayed-out sources are those falling outside the WFM FoV during a single pointing toward the Galactic Center (Background map courtesy of T. Mihara, RIKEN, JAXA, and the MAXI team).}
\label{fig:WFM_FoV}
\end{figure}

In Figure \ref{fig:WFM_FoV} we show a comparison of the WFM FoV with that of the most relevant existing facilities (background map courtesy of T. Mihara, RIKEN, JAXA, and the MAXI team). With its FoV of $1.75{\pi}$~sr at 0\% response, and $1.33{\pi}$~sr at $20$\% of the peak camera response, the WFM will have the largest FoV in comparison with that of other similar instruments. This is essential in view of the detection of counterparts of gravitational wave sources. On the map, yellow diamonds indicate known Be X-ray binary systems, red crosses indicate supergiant HMXBs, orange crosses SFXTs, and magenta boxes SyXBs. Grayed-out sources are those falling outside the WFM FoV during a single pointing toward the Galactic Center (or located in peripheral regions of the FoV where the effective area of the instrument is relatively too low to expect a reasonable sensitivity).

\begin{figure}[H]
\centering
\includegraphics[width=1\columnwidth]{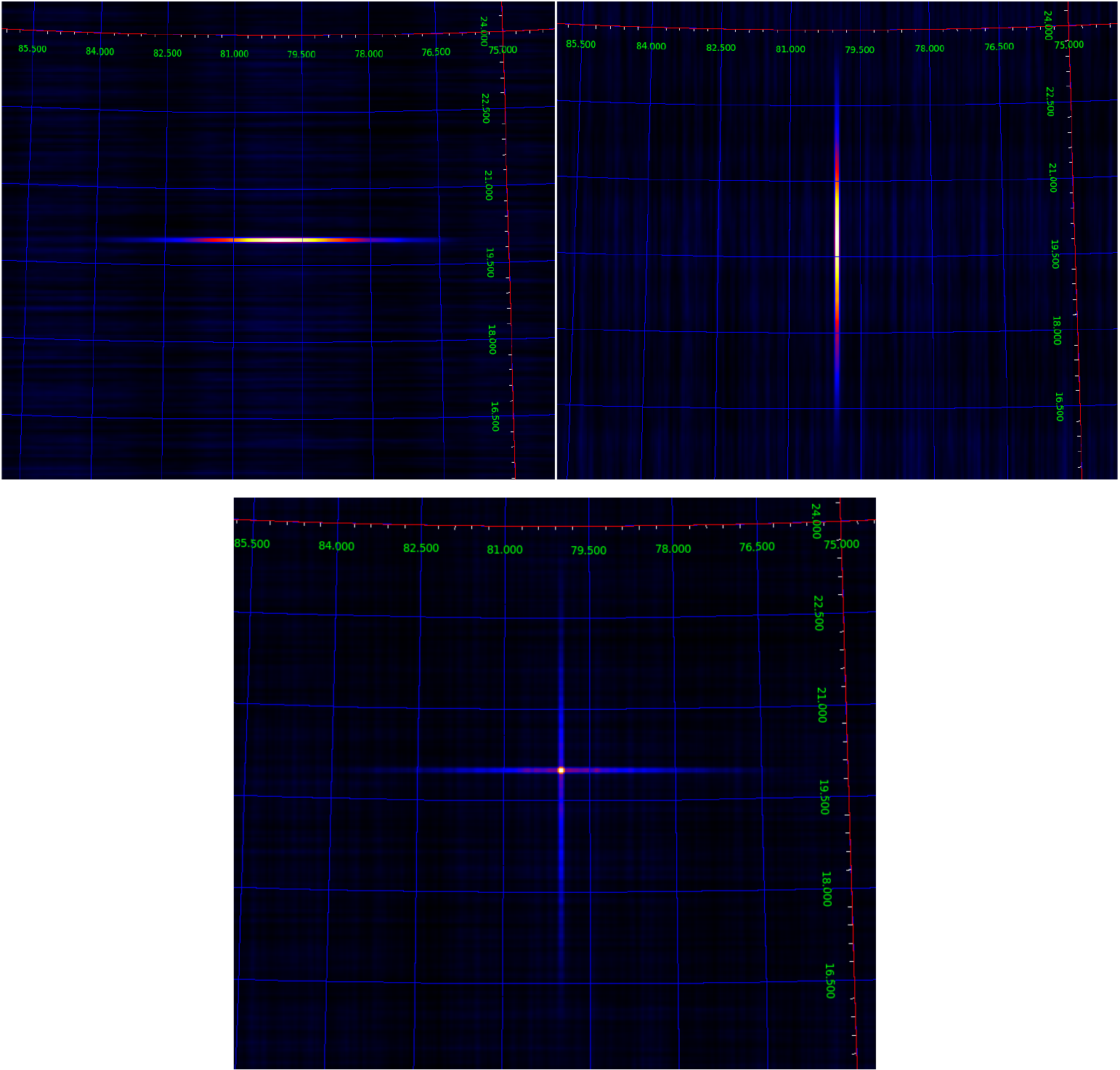}
\caption{The two-dimensional position of a source is obtained by combining the images of the two pairs of each camera, which are orthogonally oriented with respect to each other. The position accuracy in the fine direction is $<14$~arcsec (at $1 {\sigma}$).}
\label{fig:WFM-PSF}
\end{figure}

Each WFM camera produces a shadowgram convolved sky image with PSF $\sim5$~arcmin $\times 5$~deg, that often is referred to as a 1.5D image. The position accuracy in the finer direction is $<14$~arcsec (at $1 {\sigma}$). The 2D position of the source is found by combining the two independent orthogonal positions measured by the camera pairs. This is illustrated in Figure \ref{fig:WFM-PSF}.
The intensity is obtained by fitting the source strength of each camera. 

The large FoV of the WFM provides unique opportunities for detecting $\sim100$ GRBs per year. The Burst Alert System is modeled following the design of the SVOM mission and elements of the INTEGRAL BAS.  As already mentioned, the localization will drive the on-board processing power. Two onboard trigger algorithms are foreseen, an "image trigger" which performs systematic image deconvolutions on long time scales, and a "count rate trigger", which as a first first step selects short time scales of counts in excess over background to be deconvolved in a second step. The detection of an uncatalogued source gives rise to a well localized burst candidate. The on-board VHF transmitter will transmit short messages with time and sky position to a network of small ground stations heritage of SVOM. 
The goal is to deliver trigger time and burst position to end users within 30~s for fast follow-up of the fading GRB afterglow. 

\section{The mission profile}\label{sec:Mission}
The main parameters of the mission are listed in Table~\ref{tabMission} \cite{Zhanglong}.
\begin{table}[H]
\begin{center}
\caption{The main parameters of the mission.}
\label{tabMission}
\footnotesize
\begin{tabular}{l|l}
\bottomrule
\textbf{Parameter} & \textbf{Value} \\
\hline
Orbit altitude and inclination & 550~km, $<2.5^{\circ}$ \\
Launcher and launch base & LM7 + upper stage, Wenchang \\
Launch mass & 4500~kg \\
Power & $\sim$5~kW \\
Telemetry rate, Band & 3.2~Tb/day, X-band or Ka-band\\
Pointing & 3-axis stabilized, $< 0.01^{\circ}$ (3$\sigma$)\\
Ground stations& Sanya (China), Malindi (Kenya)\\
Burst Alert & VHF transmitter\\
& BeiDou Navigation Satellite System \\
Mission Lifetime & 5 years (goal 8 years) \\
Launch date & 2025\\
\bottomrule
\end{tabular}
\end{center}
\end{table}

\subsection{The spacecraft} 

\begin{figure}[H]
\centering
\includegraphics[width=0.35\columnwidth]{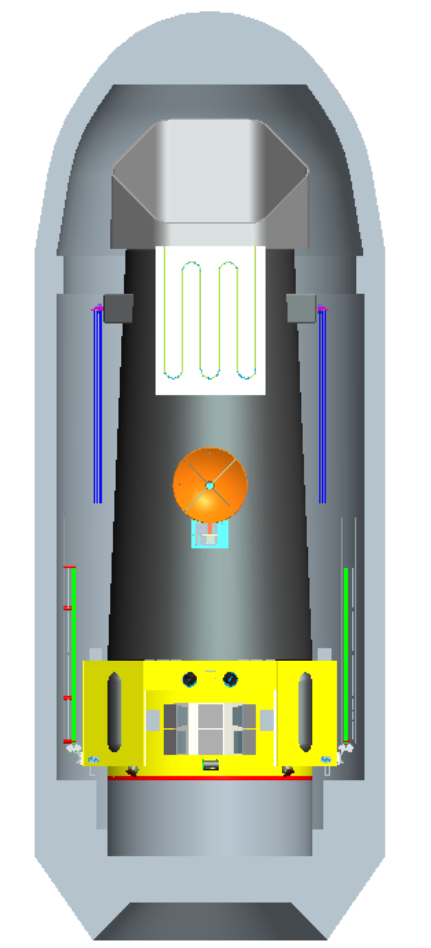}
\caption{Stowed during launch (diameter 4600 mm x 8835 mm).}
\label{fig:eXTP_stowed}
\end{figure}

\begin{figure}[H]
\centering
\includegraphics[width=1.0\columnwidth]{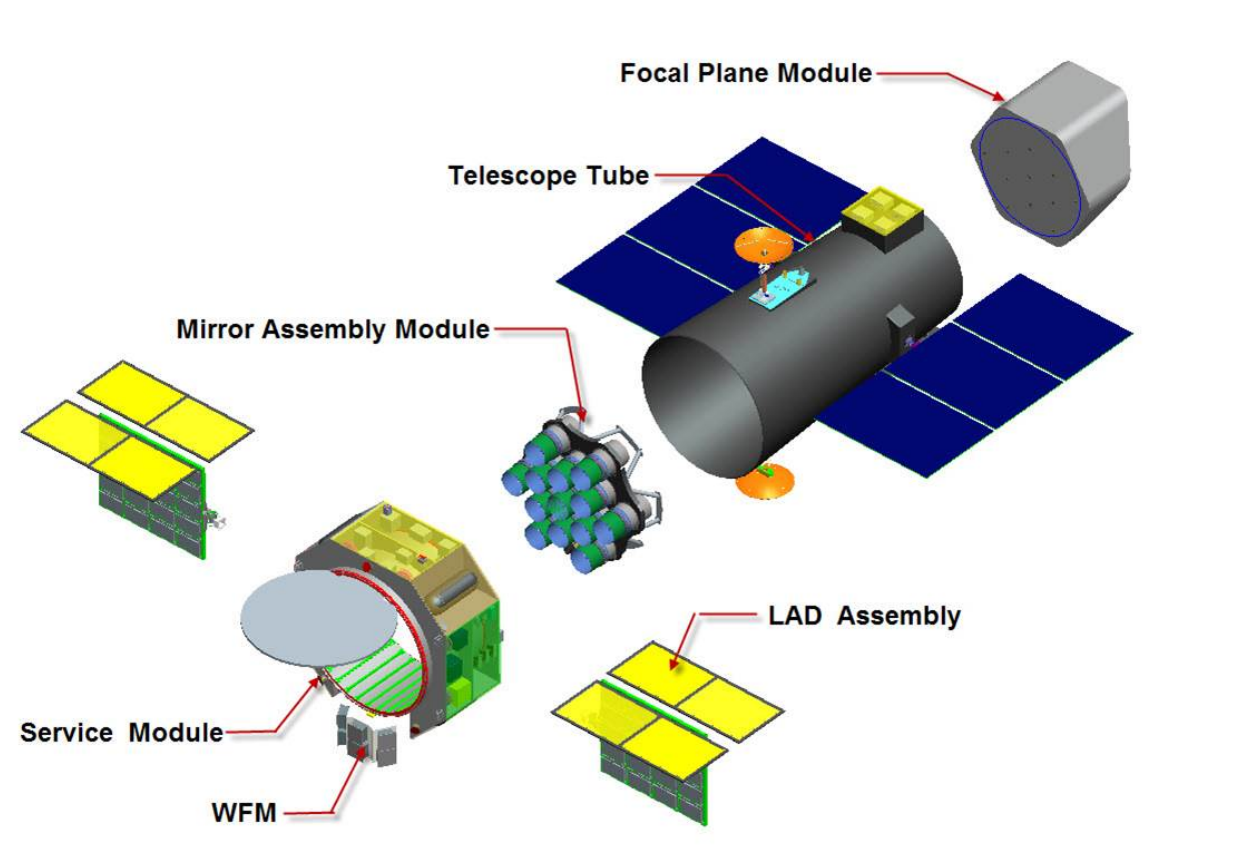}
\caption{The expanded view of the deployed satellite is shown. In orbit the eXTP encumbrance is 4737mm x 8781mm x 11500mm.}
\label{Fig:eXTP_expanded}
\end{figure}

\textit{Observation mode design and configuration}. Following the science goals, the pointing observation mode is the main operational mode of the mission. In this mode, the payload principal axis points to the target and the satellite maintains 3-axis inertial attitude stabilization. A second relevant mode is the scan mode. In this mode a small area (a few degrees) will be scanned near a known source for pointing calibration or PSF verification. The eXTP design is mainly driven by the requirement to maximize the focal length and effective area whithin the constrains inposed by the use of the LM-7 launch vehicle fairing. The mission can be divided into separated functional modules to optimize the AIT flow. The separate functional modules shown in Figure \ref{Fig:eXTP_expanded} are: the mirror assembly module (MAM), the focal plane module (FPM), the LAD Assembly (LADA), and the WFM.

The MAM consist of the SFA and PFA mirror modules and star trackers. To increase the overall observation sensitivity, the 13 mirror modules are mounted onto an optical bench, which maintains their co-alignment. The FPM hosts the 13 focal plane cameras, with sunshields to protect the detectors from solar irradiation. The telescope tube serving as the main structure of the spacecraft supports the MAM and FPM to achieve the required 5.25 m focal length. The LADA will be launched in the stowed configuration shown in Figure \ref{fig:eXTP_stowed}. After launch the two LADA wings will be deployed along the same pointing direction to the MAM.

The deployed configuration is shown in Figure \ref{Fig:eXTP_expanded}.

\textit{Subsystems design.} The mechanical system allows maintaining the required focal length without any deployment mechanisms. To meet the required structural stability, the structure is manufactured in CFRP.

The electronic system is based on the system management unit, and uses a hierarchical distributed network as the system architecture, to perform scheduling, information processing, monitoring and coordination of satellite operations, management and data processing of the science payloads, and integrated electronic systems for unified processing and sharing of satellite information. 

\subsection{Launch and Orbit}

\begin{figure}[H]
\centering
\includegraphics[width=\columnwidth]{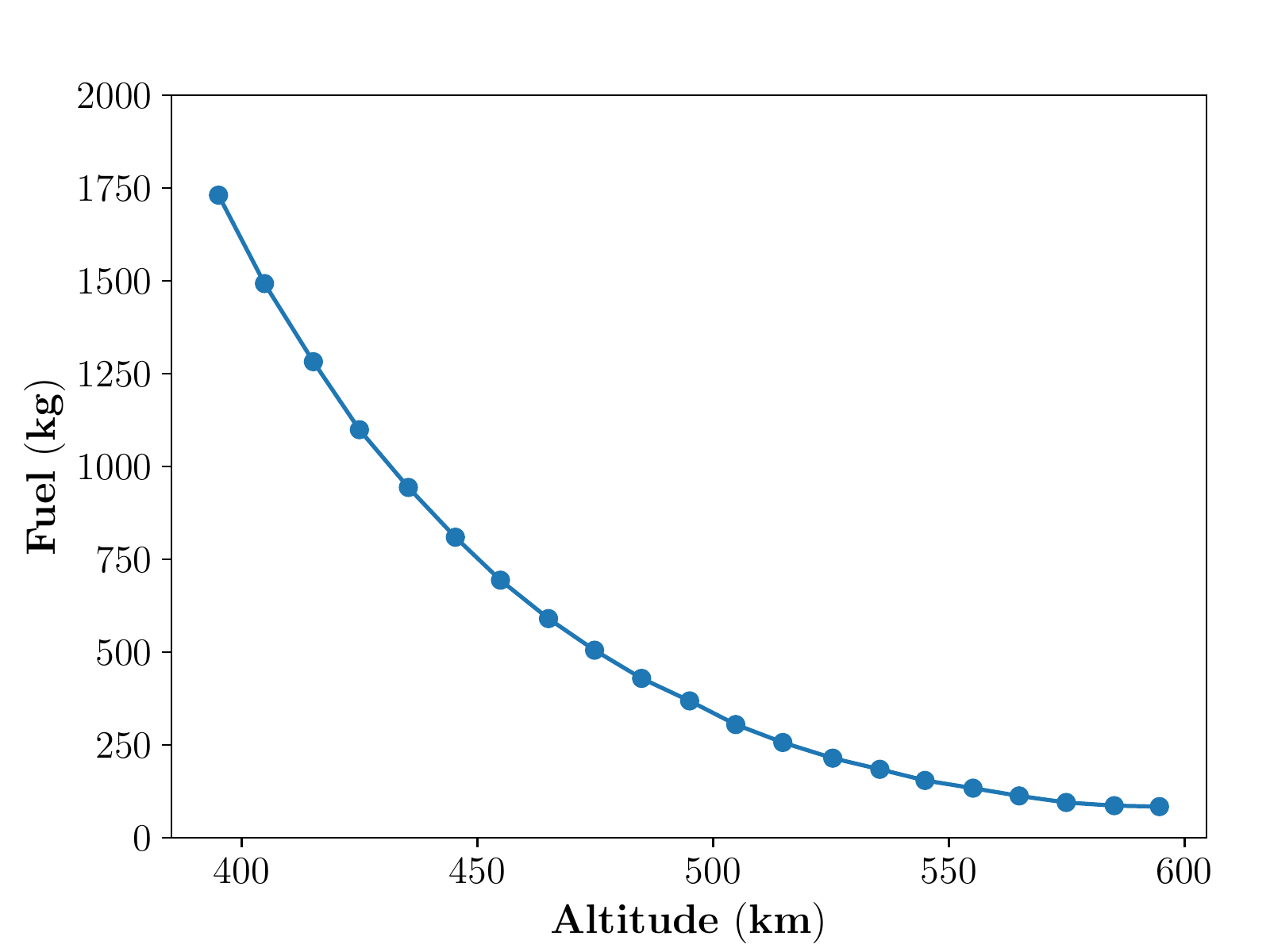}
\caption{Fuel requirements for different orbit altitudes assuming a 5 yr lifetime of the mission.}
\label{fig:Fuel_altitude}
\end{figure}

The orbit of the mission is an equatorial low earth orbit (LEO), with an inclination $<2.5^{\circ}$ and an altitude of 550~km. The orbit parameters have been selected to minimise the radiation (NIEL) damage, allowing them to be operated at temperatures compatible with passive thermal control ($-10$~$^{\circ}$C at the end of the nominal mission), with acceptable losses in the energy resolution of the instruments. In this orbit the spacecraft is robustly shielded by the geomagnetic field against solar particle events and cosmic rays. Damage on the SDDs mainly arises from trapped protons of the van Allen belts and from charge-exchange. The low altitude and inclination of the orbit minimise interactions of these particles with the spacecraft, especially reducing the through-time and therefore the effects of the South Atlantic Anomaly. A thorough study of the effects of trapped protons has been conducted for the ESA's M3 LOFT assessment phase using AP8/AP9 and Petrov radiation models (including comparisons with previous missions in LEO, e.g., BeppoSAX and Proba) \cite{DelMonte2014, Campana2014}. A low altitude and inclination orbit is also favoured based on the fuel requirements needed to keep the mission along the nominal orbit. This is shown in Figure~\ref{fig:Fuel_altitude} where results of simulated fuel consumption at different orbit altitudes are reported assuming a 5 year lifetime.

The mission is designed to be carried by the LM-7 Launch Vehicle \cite{LM7}. The envelope size of the satellite has to be within a diameter of 4600~mm, and the total height is 8835~mm, which is compatible with the constrains of the carrier fairing. The launch vehicle will place the satellite into orbit by a direct approach. According to the orbit requirements of the mission, the Wenchang Satellite Launch Center has been selected. this site has the capability of testing and launching, tracking and measuring, secure controlling, and predicting the initial orbit of the satellite. The mission TT\&C system adopts the unified S-band communication channel (USB). During the in-orbit test stage and the normal operation stage, the China Satellite Monitoring and Control Network is responsible for tracking, controlling, monitoring and uplink injection of observation commands. It has the ability to implement the tracking, telemetry and command uplink of eXTP in launching stage, in-orbit test and normal operation stage.

According to the current plans, China is responsible for the space segment, which includes the payload support, the service module and the integration of the payload. The SFA and the PFA instruments will be provided by China. Europe plans to contribute to the SFA and PFA focal plane cameras. The LAD and WFM instruments will be provided by the European part of the eXTP consortium, which fundamentally includes all countries that participated to the LOFT study. China is also responsible for the launch segment, the telemetry and control segment, and the operational and science ground segments, which will be presented in detail in the next section.
\subsection{The Ground Segment.}
\begin{figure}[H]
\centering
\includegraphics[width=1.0\columnwidth]{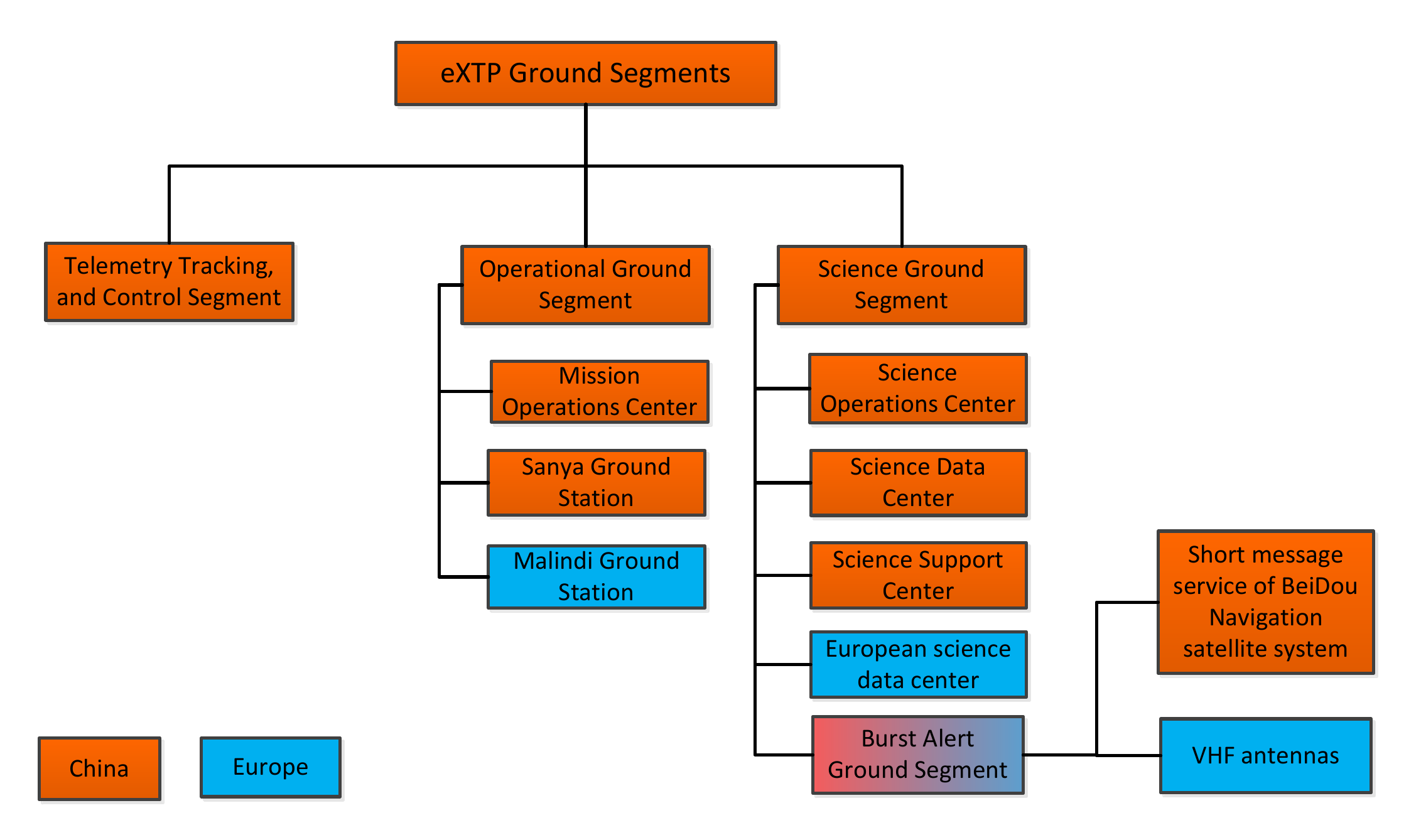}
\caption{The overall structure of the eXTP ground segments.}
\label{fig:groundsegment}
\end{figure}

The ground segment consists of three main blocks (see Figure~\ref{fig:groundsegment}): 1) the telemetry, tracking, and control segment (TT\&C); 2) the operational ground segment (OGS); 3) the science ground segment (SGS). The TT\&C leads the monitoring and verification of the spacecraft health by using the China Satellite Monitoring and Control Network, which comprises several TT\&C stations operating in the S-band. The China Satellite Monitoring and Control Network is responsible for the uplink injection of telecommands. The tracking and data relay satellite system of China is also considered for real time telemetry and control of the satellite with 100\% duty cycle while the USB system has only ten to twenty minutes contact time for each orbit. An alternative concept to distribute the burst alert message is based on the use of the VHF transmitter or the BeiDou Navigation Satellite System. The data transmission system adopts a transmission rate not lower than 2250 Mbps, and the working frequency is in Ka band (or 600-900~Mbps in X band), which allows a data transmission capacity of not less than 3.2 Tb per day.

\begin{table}[H]
\footnotesize
\caption{Main tasks of the Science Ground Segment}
\label{tab:scisegment}
\begin{tabular}{p{0.2\columnwidth} p{0.7\columnwidth}}
\toprule
  & Main tasks \\
  \hline
  Mission Operation Center (MOC) & 
  1) Receiving the raw telemetry and auxiliary files from the spacecraft; 
  2) Processing of raw telemetry into level 0 data, making them available (together with the auxiliary files) to the SDCs; 
  3) Translate the mission observational program into specific telecommands. \\ 
  \hline
  Science Operations Center (SOC) & 
  1) Schedule the observation programs, taking into account the  radiation belt passages, viewing constraints imposed by the Earth, Moon, Sun, etc.;
  2) Handle all payload anomalies;
  3) Support the SSC for the technical screening of the proposals;
  4) Monitor the status of the payload. \\
  \hline
  Chines Science Data Center (SDC) & 
  1) Develop the SFA and PFA science analysis software; 
  2) Generate the higher level scientific data products for the LAD, WFM, PFA, and SFA (level 1 to 3);
  3) Provide calibrations for the SFA and PFA;
  4) Build background models for the SFA and PFA;
  5) Monitor the performance of the payloads; 
  6) Support the SSC in the final validation of the SFA and PFA data. 
  7) Inform the community about relevant astrophysical events detected during the quick-look of the data. \\
  \hline
  European Science Data Center (SDC) &
  1) Develop the LAD and WFM science analysis software; 
  2) Provide calibrations of the LAD and WFM;
  3) Build Background models for the LAD and WFM;
  4) Provide a European data archive mirror; 
  5) Support the Chinese SDC in the monitor of the LAD and WFM performances; 
  7) Inform the community about relevant astrophysical events detected during the quick-look of the data, together with the Chinese SDC. \\
  \hline
  Science Support Center (SSC) &
  1) Provide the auxiliary software tools;
  2) Manage the data archive and the eventual Chinese mirrors; 
  3) Release the scientific data products and auxiliary data products (carrying out the final data varification);
  4) Support the scientific users;
  6) Issue the call for proposals and organize the peer review process for the proposal evaluation and selection;
  7) Receive and take decisions on ToOs and coordinated observational campaigns. \\
    \hline
  Burst Alert Ground Segment (BAGS) & 
  1) Receive, verify, and validate the on-board triggers;  
  2) Provide the required software to handle and distribute the alerts derived from the on-board triggers; 
  3) Inform the community about relevant events detected thanks to the on-board triggers. \\
\bottomrule
\end{tabular}
\end{table}
The OGS comprises the mission operation center (MOC) and the two main ground stations for the telemetry reception, i.e., the Sanya (China) and Malindi (Europe) ground stations. Both stations will be equipped with X-band receivers and transmitters. The MOC will be based at NSSC, CAS, and will be in charge of receiving the instrument telemetry, together with all relevant auxiliary files (e.g., house-keeping files). All of this data will be decoded and processed to a manageable level 0 format (typically FITS) by the MOC before making them available to the SGS for further processing, quick-look activites, archiving, and distribution. The MOC will also be in charge of translating the mission observational program provided by the SGS into specific telecommands to be delivered to the TT\&C for uplink in a transparent mode. The tasks of the main eXTP ground segment parts are shown in Table \ref{tab:scisegment}. 

The SGS, located at IHEP CAS, includes the science support center (SSC), the science operation center (SOC), and two science data centers (SDCs). The SSC is in charge of issuing calls for proposals, and organizing the peer reviewed process for their evaluation. The SSC will receive Target of Opportunity observation (ToO) requests and will help coordinating joint observational multi-wavelengths and multi-messenger programs. The SSC will also be in charge of developing the auxiliary file software tools and performing the final mission data and auxiliary file validation before making all of these products available to the scientific community through the central mission archive at NSSC, CAS. Mirrors of the archive will be hosted at different institutes in China and in Europe, to be used as backup copies. The SOC will be mainly in charge of building up the yearly mission observational plan, and of monitoring the status and performance of the payload. The Chinese and European SDCs will provide the analysis software to process and handle data from the scientific payload, and pre-defined science products. These products will be used to perform quick-look activities. 
According to the current plans, the WFM data will be made publicly available as soon as they are processed on the ground. 
The on-board burst alert systems will transmit alerts from bright impulsive events to the ground and the science community within $\lesssim$30~s. 15 VHF ground stations deployed around the equator will be used to download the burst alert system information. These stations are designed according to the SVOM heritage and deployed in a sub-set of their sites (or pre-existing stations from SVOM). Alternatively, the short message service of BeiDou Navigation Satellite System or relay satellite communication system could also be used. 
\section{Observing strategy}\label{sec:4}
The prime goal of the observing strategy is to ensure the best possible scientific results for the mission, opening the mission to the scientific community, while ensuring a return to the institutions and countries that financially supported the mission. It is therefore foreseen that eXTP will operate as an observatory open to the scientific community. Observing time will be allocated via annual Announcement of Opportunities and through scientific peer review. A fraction of the observing time will be allocated to key core projects that should ensure that the core science goals of the mission are reached. According to the current estimates, a total of $\sim$45 Ms shall be allocated to  key projects. The mission will be designed to fast react to transients events, e.g., outbursts, performing flexible ToOs. ToOs (outside the state changes of sources defined in the guest observer program and in the key projects) can be proposed by the scientific community and the eXTP PI will decide whether to schedule the proposed observation, after consulting the relevant members of the OTAC. Depending on the nature of the ToO either the one year proprietary data rule can be applied or the data will be made available to the science community in general. A standard one year proprietary data rights are assumed for the pointed instruments, and after this period all data shall become public. For the WFM no proprietary data rights will apply and this data should be made available to the community on a much shorter time scale. During routine operations several pre-products data will be produced typically in 3 hours after the observation (NRT data). The data of the Burst Alert System will be distributed to the community in typically 30 seconds using the appropriate systems and no data rights will apply.

\section{Conclusions}\label{sec:5}
In this paper we have presented the tecnical and technological aspects of the scientific payload of the eXTP mission, the performance of the instruments, and the main aspects of the mission. The eXTP mission is designed to address key questions of physics in the extreme conditions of ultradense matter, strong field gravity, and the strongest magnetic field existing in nature. The payload includes three narrow field instruments: the SFA, characterized by large area at soft energies, the LAD, with unprecedented area at hard X-rays, and the PFA, a polarimeter between five and six times more sensitive than the polarimetr onboard the NASA mission IXPE. The science payload is completed by wide field monitoring capabilities. This combination will enable for the first time ever spectral-timing-polarimetry studies of a large number of objects populating the time variable Universe. In addition to investigating fundamental physics, eXTP will be a very powerful observatory for astrophysics, providing data on a variety of galactic and extragalactic objects. In particular, the WFM will be highly instrumental to detect the electro-magnetic counterparts of transients sources of the multi-messenger Universe including gravitational wave sources, and cosmic neutrinos sources

The mission is led by China, currently supported jointly by CAS and China National Space Administration, and executed under the management of the National Space Science Center of CAS, in the framework of the activities of the CAS Bureau of Major R\&D Programs. IHEP of CAS leads the science consortium and is responsible for coordinating
all science payload contributions within China and from the international partner countries and agencies.
The extended phase A study of the mission will be completed at the end of 2018. Very recently, NSSC, CAS, and the ESA's science directorate have started a joint study to identify the potential contribution of ESA to the mission through a mission of opportunity. The Italian space agency ASI has also initiated the coordination of the expectd contributions from the European partners. Provided the successfull conclusion of the extended phase A, phase B will be carried out in 2019. Mission adoption is foreseen in December of 2019 or January 2020. According to the current schedule phase C and D will be carried out in 2020-2024 so to achieve the launch of the mission in 2025.


\emph{Acknowledgements.} The  Chinese  team  acknowledges  the  support  of  the  Chinese  Academy  of  Sciences  through  the  Strategic  Priority Research Program of the Chinese  Academy of Sciences, Grant No. XDA15020100. The Italian collaboration acknowledges support by ASI, under the dedicated eXTP agreements and agreement ASI-INAF n.2017-14-H.O., by INAF and INFN under project REDSOX. The German team acknowledges support from the Deutsche Zentrum f{\"u}r Luft- und Raumfahrt, the German Aerospce Center (DLR). The Polish Team acknowledges the support of Science Centre grant 2013/10/M/ST9/00729. The Spanish authors acknowledge support from MINECO grant ESP2017-82674-R and FEDER funds.













\end{multicols}
\end{document}